\documentclass[]{jfm}
\usepackage{natbib}
\usepackage{amsmath,bm}
\usepackage[dvips]{graphicx}
\usepackage{color}
\usepackage{tikz}
\usetikzlibrary{arrows,decorations.markings}
\usepackage{tikz-3dplot}
\usepackage{pgfplots}

\newcommand{\aver}[1]{ \! \left\langle {#1} \right \rangle \!}

\newcommand{\linesample}[2]{\raisebox{2pt}{\tikz{\draw[{#1},{#2},thick](0,0)--(6mm,0);}}}

\newcommand{\deriv}[2]{\frac{\partial {#1}}{\partial {#2}}}
\newcommand{\Deriv}[2]{\frac{\mathrm{d} {#1}}{\mathrm{d} {#2}}}
\newcommand{\phiR}{\phi}
\newcommand{\phir}[2]{\phiR_{{#1},{#2}}}
\newcommand{\phiC}{\psi}
\newcommand{\phic}[3][,]{\phiC_{{#2}{#1}{#3}}}
\newcommand{\src}[1][]{\xi_{{#1}}}
\newcommand{\Xc}{X}
\newcommand{\R}{r}

\newcommand{\del}[1][u]{\delta {#1}}
\newcommand{\duiduj}{\del_i \del_j}

\newcommand{\kron}{\delta}
\newcommand{\var}{V}
\renewcommand{\mid}[1]{{#1}^{\ast}}

\begin{document}
\title[Structure function tensor equations in inhomogeneous turbulence]
{Structure function tensor equations \\ in inhomogeneous turbulence}

\author[D.Gatti, A.Chiarini, A.Cimarelli \& M.Quadrio]{
D\ls A\ls V\ls I\ls D\ls E\ns G\ls A\ls T\ls T\ls I$^1$ \thanks{Email address for correspondence: davide.gatti@kit.edu}, \ls
A\ls L\ls E\ls S\ls S\ls A\ls N\ls D\ls R\ls O\ns C\ls H\ls I\ls A\ls R\ls I\ls N\ls I$^2$, \ls
A\ls N\ls D\ls R\ls E\ls A\ns C\ls I\ls M\ls A\ls R\ls E\ls L\ls L\ls I$^3$ \ls
\and \ns
M\ls A\ls U\ls R\ls I\ls Z\ls I\ls O\ls \ns Q\ls U\ls A\ls D\ls R\ls I\ls O$^2$
\thanks{Mercator Fellow at Karlsruhe Institute of Technology}
}
\affiliation{
$^1$Institute of Fluid Mechanics, Karlsruhe Institute of Technology,
Kaiserstra\ss e 10, 76131 Karlsruhe, Germany
\\[\affilskip]
$^2$Department of Aerospace Sciences and Technologies, Politecnico di Milano,
via La Masa 34, 20156 Milano, Italy
\\[\affilskip]
$^3$Dipartimento di Ingegneria ``Enzo Ferrari'', Universit\`a di Modena e Reggio Emilia, 41125 Modena, Italy
}

\date{\today}

\maketitle

\begin{abstract}
Exact budget equations for the second-order structure function tensor $\aver{\delta u_i \delta u_j}$ are used to study the two-point statistics of velocity fluctuations in inhomogeneous turbulence. The Anisotropic Generalized Kolmogorov Equations (AGKE) describe the production, transport, redistribution and dissipation of every Reynolds stress component occurring simultaneously among different scales and in space, i.e. along directions of statistical inhomogeneity. The AGKE are effective to study the inter-component and multi-scale processes of turbulence. In contrast to more classic approaches, such as those based on the spectral decomposition of the velocity field, the AGKE provide a natural definition of scales in the inhomogeneous directions, and describe fluxes across such scales too. Compared to the Generalized Kolmogorov Equation, which is recovered as their half trace, the AGKE can describe inter-component energy transfers occurring via the pressure-strain term and contain also budget equations for the off-diagonal components of $\aver{\delta u_i \delta u_j}$. 

The non-trivial physical interpretation of the AGKE terms is demonstrated with three examples. First, the near-wall cycle of a turbulent channel flow at $Re_\tau=200$ is considered. The off-diagonal component $\aver{-\delta u \delta v}$,  which can not be interpreted in terms of scale energy, is discussed in detail. Wall-normal scales in the outer turbulence cycle are then discussed by applying the AGKE to channel flows at $Re_\tau=500$ and $1000$. In a third example, the AGKE are computed for a separating and reattaching flow. The process of spanwise-vortex formation in the reverse boundary layer within the separation bubble is discussed for the first time.
\end{abstract}

\section{Introduction}
\label{sec:intro}

Since the early days of fluid mechanics, understanding turbulence fascinates scholars, enticed by the goal of identifying the key mechanisms governing turbulent fluctuations and eventually determining the mean flow. This is essential for developing and improving RANS and LES turbulence models, useful in engineering practice. Most turbulent flows of applicative interest, in particular,  are challenging because of their anisotropic and inhomogeneous nature.

Among the several approaches pursued so far to address the physics of inhomogeneous and anisotropic turbulence, the two most common ones observe the flow either in the space of scales, or in the physical space. In the scale-space approach, the characteristic shape and size of the statistically most significant structures of turbulence are deduced from two-point second-order statistics. A spectral decomposition of the velocity field can be employed to describe the scale distribution of energy, while spatial correlation functions are used to characterise the shape of the so-called coherent structures \citep{robinson-1991b, jimenez-2018}. Since a turbulent flow contains eddies of different scales, the power spectral density of turbulent fluctuations is a gauge to the actual eddy population, and provides useful information to develop kinematic models of turbulence capable to explain some of its features. One such model rests on the attached-eddy hypothesis by \cite{townsend-1976}, and predicts self-similar features of turbulent spectra in wall-bounded flows \citep{perry-chong-1982}. Two-points correlations of velocity fluctuations are the inverse Fourier transform of power spectra. They emphasise the spatial coherence of the largest and strongest turbulent fluctuations, and have been, for instance, employed to describe the streaky structure of near-wall turbulence \citep{kline-etal-1967}, to identify large-scale structures in high-Reynolds number flows \citep{smits-mckeon-marusic-2011,sillero-jimenez-moser-2014} or to describe the structural properties of highly-inhomogeneous separating and reattaching turbulent flows \citep{mollicone-etal-2018, cimarelli-leonforte-angeli-2018}.

In the physical-space approach, it is possible to characterise the spatial organisation of production, transfer and dissipation of the turbulent kinetic energy associated with the temporal fluctuations of the three velocity components. The tools of choice are the exact single-point budget equations for the components of the Reynolds stress tensor and of its half-trace, the turbulent kinetic energy $k$. 
This approach has been successfully applied to canonical wall-bounded flows and, more recently, to more complex turbulent flows. For the former, the main focus has been the inhomogeneity and anisotropy induced by the wall \citep{mansour-kim-moin-1988} and the effect of the Reynolds number \citep{hoyas-jimenez-2008} on the Reynolds stress budgets. For the latter, the Reynolds stress production and transport phenomena have been studied in free shear layers and recirculation bubbles  \citep{mollicone-etal-2017, cimarelli-leonforte-angeli-2018, cimarelli-etal-2019-resolved}, where local non-equilibrium results in significantly different physics.

\begin{figure}
\centering
\includegraphics[width=\textwidth]{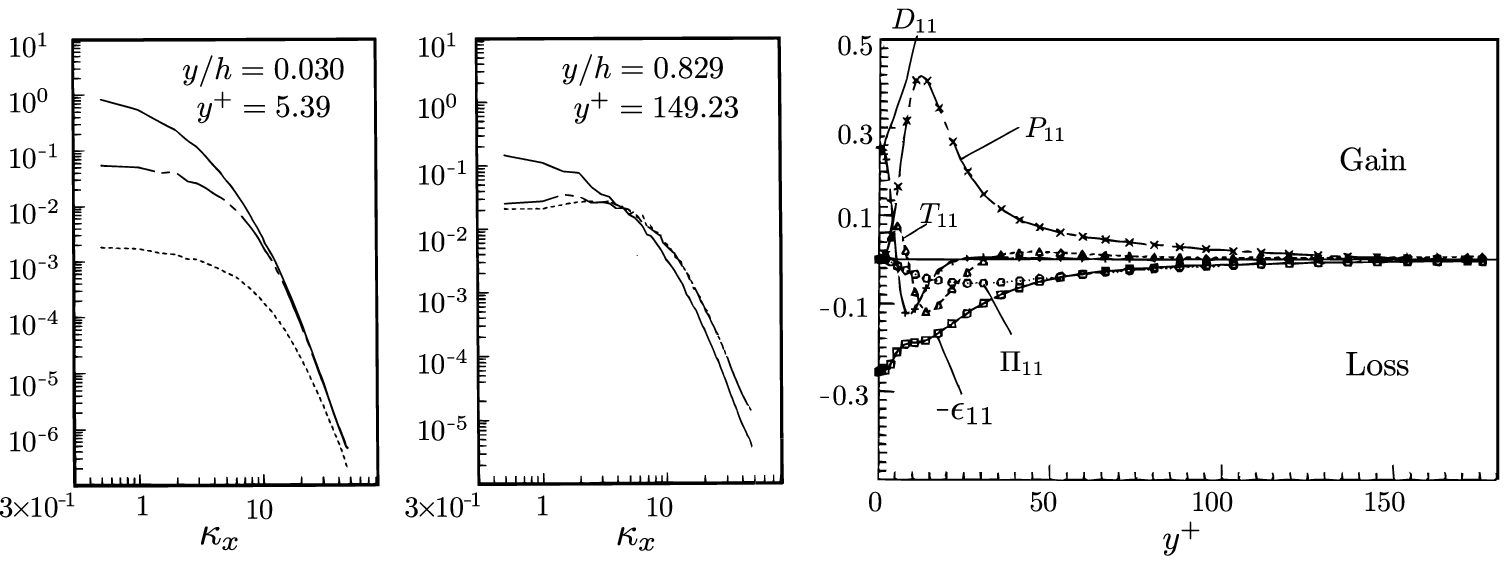}
\caption{Second-order statistics after the seminal DNS of a turbulent channel flow by \cite{kim-moin-moser-1987}. Left, adapted from \cite{kim-moin-moser-1987}: one-dimensional energy spectra versus streamwise wavenumber $\kappa_x$, at two wall distances. Continuous, dashed and dotted lines refer to streamwise, spanwise and wall-normal velocity fluctuations.  Right, adapted from \cite{mansour-kim-moin-1988}: terms in the budget equation for $\aver{u'_1 u'_1}$, with notation as in the original paper. $P_{11}$: production; $\epsilon_{11}$: dissipation; $\Pi_{11}$: velocity pressure-gradient term; $T_{11}$: turbulent transport; $D_{11}$: viscous diffusion.}
\label{fig:example}
\end{figure}

Typical results ensuing from the two approaches above are exemplified in figure \ref{fig:example}, where key plots from \cite{kim-moin-moser-1987} and from \cite{mansour-kim-moin-1988} are reproduced. Both diagrams stem from the analysis of the same DNS database for a low-$Re$ turbulent channel flow. The two leftmost plots are one-dimensional turbulent energy spectra as function of the streamwise wavenumber, each computed at a specific distance from the wall. The right plot shows the wall-normal behaviour of the terms appearing in the budget of the 1,1 component of the Reynolds stress tensor.

Despite their fundamental importance, both approaches suffer of some limitations. Indeed, it is well known since \cite{richardson-1922} that turbulence is a truly multi-scale phenomenon, where fluctuations of different spatial extent non-linearly interact through energy-cascading mechanisms. Even more so, in inhomogeneous  flows these interactions vary in space significantly, leading to 
a transfer of momentum between different spatial locations. 
The single-point budget equations for the Reynolds stresses do not contain information about the scales involved in such energy fluxes, and therefore miss the multi-scale nature of turbulence. 
The spectral decomposition and two-point spatial correlations do discern the different scales, but fail to provide direct information on their role in the processes of production, transfer and dissipation of $k$, and therefore lack a dynamical description of turbulent interactions.

These limitations are overcome when space and scale properties of turbulence are considered {\em jointly}. For example, to recover the scale information \cite{lumley-1964}, \cite{domaradzki-etal-1994} and more recently \cite{mizuno-2016} and \cite{lee-moser-2019} analysed spectrally decomposed budget equations for the Reynolds stresses. They observed inverse energy transfers from small to large scales, supporting substantial modifications of the Richardson scenario in wall-bounded flows. Unfortunately, however, spectral analysis does not allow a definition of scales in statistically inhomogeneous directions, such as the wall-normal one in wall-bounded flows. \cite{hill-2001}, \cite{danaila-etal-2001}, \cite{hill-2002} and \cite{dubrulle-2019} proposed a complementary approach, free from this restriction, and generalized the \cite{kolmogorov-1941} description of the energy transfer among scales from isotropic flows to inhomogeneous flows. 

The Generalized Kolmogorov Equation or GKE \citep[see for example][]{danaila-antonia-burattini-2004, marati-casciola-piva-2004, rincon-2006, cimarelli-deangelis-casciola-2013, cimarelli-etal-2015, cimarelli-etal-2016, portela-papadakis-vassilicos-2017} is an exact  budget equation for the trace of the so-called second-order structure function tensor, i.e. the sum of the squared increments in all three velocity components between two points in space. This quantity is interpreted as scale energy, and provides scale and space information in every spatial direction, regardless of its statistical homogeneity. The present work discusses the Anisotropic Generalized Kolmogorov Equations (AGKE), which extend the scale and space description of the GKE, limited to scale energy. The goal is to describe each component of the structure function tensor separately, thus capturing the anisotropy of the Reynolds stress tensor and of the underlying budget equations. This provides a complete description of energy redistribution among the various Reynolds stresses. The AGKE identify scales and regions of the flow involved in the production, transfer and dissipation of turbulent stresses, thus integrating the dynamical picture provided by single-point Reynolds stress budgets with the scale information provided by the spectral decomposition. The relationship between the second-order velocity increments and the two-point spatial correlation functions can be exploited to identify the topological features of the structures involved in creation, transport and destruction of turbulent stresses. This endows the kinematic information provided by the spatial correlation functions with additional dynamical information from exact budget equations. 

The present work aims at introducing the reader to the AGKE and to their use via example applications to inhomogeneous turbulent flows. The paper is structured as follows. First, in \S\ref{sec:AGKE} the budget equations for the structure function tensor are presented and provided with a physical interpretation, and the numerical datasets used in the example flows are described in \S\ref{sec:DNS}. 
Then AGKE are applied to canonical turbulent channel flows. In particular, \S\ref{sec:inner} focuses on the near-wall turbulence cycle of a low-$Re$ channel flow. The energy exchange among the diagonal terms of the structure function tensor via the pressure-strain term is discussed, and the complete AGKE budget of the off-diagonal component is described for the first time. Then, \S\ref{sec:outer} demonstrates the capability of the AGKE to disentangle the dynamics of flows with a broader range of scales by considering  the outer cycle of wall-turbulence in channel flows at higher Reynolds numbers. Finally, \S\ref{sec:barc} considers the separating and reattaching flow over a finite rectangular cylinder, and shows how the AGKE do in such highly inhomogeneous flows. The paper is closed by a brief discussion in \S\ref{sec:conclusions}. Additional material is reported in three appendices. The complete derivation of the AGKE and their complete form, both in tensorial and component-wise notation, are detailed for reference in Appendix A. Appendix B lists the symmetries of the AGKE terms in the specialised form valid for the indefinite plane channel. Appendix C describes the computation of the velocity field induced by the ensemble-averaged quasi-streamwise vortex, employed in \S\ref{sec:inner}.

\section{Anisotropic Generalized Kolmogorov Equations (AGKE)}
\label{sec:AGKE}

\begin{figure}
\centering
\includegraphics[]{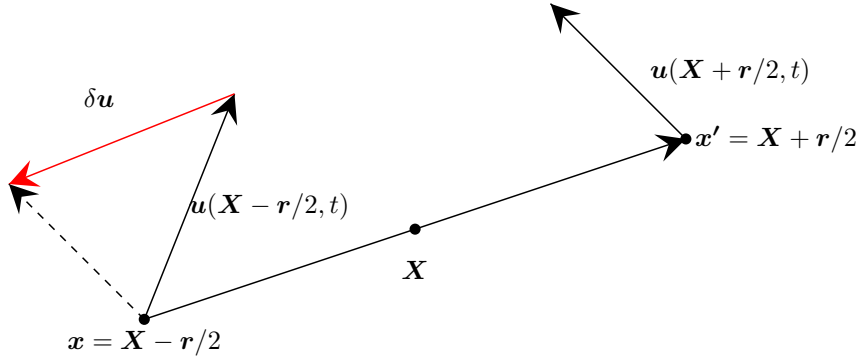}
\caption{Sketch of the quantities involved in the definition of the second-order structure function. $\bm{x}=\bm{\Xc}-\bm{\R}/2$ and $\bm{x^\prime}=\bm{\Xc}+\bm{\R}/2$ are the two points across which the velocity increment $\delta \bm{u}$ is computed.} 
\label{fig:sf-sketch}
\end{figure}

Let us consider an incompressible turbulent flow, described via its mean and fluctuating velocity fields, $U_i$ and $u_i$ respectively, defined after  Reynolds decomposition. The Anisotropic Generalized Kolmogorov Equations or AGKE are exact budget equations for the second-order structure function tensor $\aver{\duiduj}$, derived from the Navier--Stokes equations. The operator $\aver{\cdot}$ denotes ensemble averaging, as well as averaging along  homogeneous directions, if available, and over time if the flow is statistically stationary. The structure function tensor features the velocity increment $\del_i$ of the $i$-th velocity component between two points $\bm{x}$ and $\bm{x^\prime}$ identified by their midpoint $\bm{\Xc}=\left(\bm{x} + \bm{x^\prime} \right) / 2$ and separation $\bm{\R}=\bm{x^\prime} - \bm{x}$, i.e. $\del_i = u_i\left(\bm{\Xc} +\bm{\R}/2,t \right) - u_i\left(\bm{\Xc} -\bm{\R}/2,t \right)$. (In the following, unless index notation is used, vectors are indicated in bold.)

In the general case, $\aver{\duiduj}$ depends upon seven independent variables, i.e. the six coordinates of the vectors $\bm{\Xc}$ and $\bm{\R}$ and time $t$, as schematically shown in figure \ref{fig:sf-sketch}, and 
is related \citep{davidson-nickels-krogstad-2006, agostini-leschziner-2017} to the variance of the velocity fluctuations (i.e. the Reynolds stresses) and the spatial cross-correlation function as follows:
\begin{equation}
\centering
\aver{\duiduj}(\bm{\Xc},\bm{\R},t)= \var_{ij}(\bm{\Xc},\bm{\R},t) - R_{ij}(\bm{\Xc},\bm{\R},t) - R_{ij}(\bm{\Xc},-\bm{\R},t)
\label{eq:sf-corr}
\end{equation}
where
\begin{align}
\var_{ij}(\bm{\Xc},\bm{\R},t)=\aver{u_i u_j}(\bm{\Xc}+\frac{\bm{\R}}{2},t)+ \aver{u_i u_j}(\bm{\Xc}-\frac{\bm{\R}}{2},t)
\end{align}
is the sum of the single-point Reynolds stresses evaluated at the two points $\bm{\Xc}+\bm{\R}/2$ and $\bm{\Xc}-\bm{\R}/2$ at time $t$, and
\begin{align}
R_{ij}(\bm{\Xc},\bm{\R},t)= \aver{u_i\left( \bm{\Xc}+\frac{\bm{\R}}{2},t \right) u_j \left( \bm{\Xc}-\frac{\bm{\R}}{2},t  \right)}
\end{align}
is the two-point spatial cross-correlation function. The AGKE contains the structural information of $R_{ij}$; however, for large enough $\left| \bm{\R} \right|$  the correlation vanishes, and $\aver{\duiduj}$ reduces to $\var_{ij}$, whereas the AGKE become the sum of the single-point Reynolds stress budgets at $\bm{\Xc} \pm \bm{\R}/2$.

\subsection{Budget equations}

The budget equations for $\aver{\duiduj}$ describe production, transport and dissipation of the turbulent stresses in the compound space of scales and positions, and fully account for the anisotropy of turbulence. For a statistically unsteady turbulent flow, these equations link the variation in time of $\aver{\duiduj}$ at a given scale and position, to the instantaneous unbalance among production, inter-component transfer, transport and dissipation. The full derivation starting from the Navier--Stokes equations is detailed in Appendix \ref{sec:derivation}, and Appendix \ref{sec:symmetries} mentions the symmetries that apply in the plane channel case.

The AGKE can be cast in the following compact form (repeated indices imply summation):
\begin{equation}
\deriv{\aver{\duiduj}}{t} + \deriv{\phir{k}{ij}}{\R_k}  + \deriv{\phic{k}{ij}}{\Xc_k} = \src[ij] \,.
\label{eq:AGKE_gen}
\end{equation} 
For each $(i,j)$ pair, $\phir{k}{ij}$ and $\phic{k}{ij}$ are the components in the space of scales $\R_k$ and in the physical space $\Xc_k$ of a six-dimensional vector field of fluxes $\bm{\Phi}_{ij}$, and are given by:
\begin{equation}
\phir{k}{ij} =\underbrace{\aver{\del[U_k] \duiduj } }_{\text{mean transport}} +  
              \underbrace{\aver{\del[u_k] \duiduj} }_{\text{turbulent transport}}  
              \underbrace{ -2 \nu \deriv{}{\R_k} \aver{\duiduj}}_{\text{viscous diffusion}} \ \ \ k=1,2,3
\label{eq:phi_r_gen}
\end{equation}
\begin{equation}
\phic{k}{ij} =  \underbrace{ \aver{\mid{U_k} \duiduj} }_{\text{mean transport}}+
                \underbrace{ \aver{\mid{u_k} \duiduj} }_{\text{turbulent transport}} + 
                \underbrace{\frac{1}{\rho} \aver{\del[p] \del_i} \kron_{kj} + \frac{1}{\rho} \aver{\del[p] \del_j} \kron_{ki}}_{\text{pressure transport}}
                \underbrace{- \frac{\nu}{2} \deriv{}{\Xc_k} \aver{\duiduj}}_{\text{viscous diffusion}}  \ \ \ k=1,2,3
\label{eq:phi_c_gen}
\end{equation}
and $\xi_{ij}$ is the source term for $\aver{ \delta u_i \delta u_j}$:
\begin{equation}
\begin{split}
\src[ij] =& \underbrace{-\aver{\mid{u_k} \del_j} \del[\left( \deriv{U_i}{x_k} \right)] - \aver{\mid{u_k} \del_i} \del[\left(\deriv{U_j}{x_k}\right)]}_{\text{production } (P_{ij})} +\\
          & \underbrace{-\aver{\del_k \del_j} \mid{\left(\deriv{U_i}{x_k}\right)} - \aver{\del_k \del_i} \mid{\left( \deriv{U_j}{x_k} \right)}}_{\text{production } (P_{ij})} +\\
          & \underbrace{+\frac{1}{\rho}\aver{\del[p] \deriv{\del_i}{\Xc_j}} + \frac{1}{\rho}\aver{\del[p] \deriv{\del_j}{\Xc_i}}}_{\text{pressure strain } (\Pi_{ij})} %
            \underbrace{-4 \mid{\epsilon_{ij}} }_{\text{ps.dissipation } (D_{ij})} \,.
\end{split}
\label{eq:xi_gen}
\end{equation}
Here $\kron_{ij}$ is the Kronecker delta, $\nu$ is the kinematic viscosity, the asterisk superscript $\mid{f}$ denotes the average of the generic quantity $f$ between positions $\bm{\Xc} \pm \bm{\R}/2$, and $\epsilon_{ij}$ is the pseudo-dissipation tensor, whose trace is the pseudo-dissipation $\epsilon$. The sum of the equations for the three diagonal components of $\aver{ \duiduj }$ reduces to the Generalized Kolmogorov Equation \citep{hill-2001}.

Each term contributing to the fluxes in equations \eqref{eq:phi_r_gen} and \eqref{eq:phi_c_gen} can be readily interpreted in analogy with the single-point budget equation for the Reynolds stresses \citep[see e.g.][]{pope-2000} as the mean and turbulent transport, pressure transport and viscous diffusion. $\bm{\phiR}_{ij}$ describes the flux of $\aver{\duiduj}$ among scales, and turbulent transport is the sole nonlinear term. $\bm{\phiC}_{ij}$ describes the flux of $\aver{\duiduj}$ in physical space, and all its terms but the viscous one are nonlinear. The source term $\src[ij]$ describes the net production of $\aver{\duiduj}$ in space and among scales; it is similar to the one appearing in the GKE, but additionally features a pressure-strain term, involved in the energy redistribution process between different components of turbulent stresses. Each term in equation \eqref{eq:AGKE_gen} informs on the spatial position $\bm{\Xc}$, scale $\bm{\R}$ and time $t$ at which production, transport and dissipation of Reynolds stresses are statistically important.

The diagonal components of $\aver{\duiduj}$ are positive by definition, and their budget equations inherit the interpretation proposed by \cite{marati-casciola-piva-2004} and \cite{cimarelli-deangelis-casciola-2013} for the GKE: they are analogous to scale energy, and the AGKE enables their discrimination into the separate diagonal components of the Reynolds stress tensor. The non-diagonal components, however, can in general assume  positive or negative values, also when the sign of $\aver{u_i u_j}$ can be predicted on physical grounds. For these components, $\src[ij]$ has the generic meaning of a source term, which can be viewed as production or dissipation only upon considering the actual sign of $\aver{\duiduj}$ at the particular values of $\left(\bm{\Xc}, \bm{\R}\right)$. In analogy with the concept of energy cascade, paths of $\aver{\duiduj}$ in the $(\bm{\Xc},\bm{\R})$ space represent fluxes of Reynolds stresses through space ($\bm{\Xc}$) and scales ($\bm{\R}$) at time $t$. The shape of the paths is determined by $\bm{\phiC}_{ij}$ (space fluxes) and $\bm{\phiR}_{ij}$ (scale fluxes).

\subsection{Simulations and databases}
\label{sec:DNS}

\begin{table}
\begin{tabular*}{1.0\textwidth}{@{\extracolsep{\fill}} l c c c c c c c c}
$Re_\tau$ & $10^3 C_f$ & $\left( L_x ,  L_z \right)/h$ & $ N_x ,  N_y , N_z $ & $\Delta x^+$ & $\Delta z^+$ & $\Delta y^+_{\mathrm{min}}$ & $N$ &  $\Delta t \,u_\tau /h$\\[10pt]
200 &  7.93    & $4\pi , 2\pi$  &  $256 , 256 , 256$  & 6.5 & 3.3 & 0.46  & 200 & 0.62  \\
500 &  6.05    & $4\pi , 2\pi$  &  $512 , 250 , 512$  & 8.2 & 4.1 & 0.96  & 38  &  1.00 \\	
1000 &  5.00  & $4\pi , 2\pi$  &  $1024, 500 , 1024$ & 8.2 & 4.1 & 0.96  & 38  &  0.60  \\	
\end{tabular*}
\caption{Details of the three turbulent channel flow DNS databases. For each $Re_\tau$, the table provides the computed value of the friction coefficient $C_f=2(u_\tau/U_b)^2$, the size of the computational domain, number of Fourier modes and collocation points in the wall-normal direction, spatial resolution (computed after the 3/2-rule dealiasing in the homogeneous directions), the number $N$ of accumulated flow snapshots and their temporal spacing $\Delta t$. The cases at $Re_\tau=200$ and $Re_\tau=1000$ were already documented by \cite{gatti-quadrio-2016} and \cite{gatti-etal-2018}.}
\label{tab:discretization-parameters}
\end{table}

As anticipated in \S\ref{sec:intro}, the AGKE analysis below stems from the post-processing of velocity and pressure fields obtained via Direct Numerical Simulations (DNS) of two flows. The former is the turbulent plane channel flow, whose inner and outer turbulent cycles will be discussed in \S\ref{sec:inner} and \S\ref{sec:outer} respectively. The latter is the separating and reattaching flow around a finite rectangular cylinder, discussed in \S\ref{sec:barc}. 

The turbulent channel flow simulations have been carried out for the present work via the DNS code introduced by \cite{luchini-quadrio-2006}. The incompressible Navier--Stokes equations are projected in the divergence-free space of the wall-normal components of the velocity and vorticity vectors and solved by means of a pseudo-spectral method, as in \cite{kim-moin-moser-1987}. Three database are used, with friction Reynolds number $Re_\tau = u_\tau h /\nu$ of $Re_\tau=200$, $500$ and $1000$. Here $h$ is the channel half-height, and $u_\tau =\sqrt{\tau_w/\rho}$ is the friction velocity expressed in terms of the average wall shear stress $ \tau_w $ and the density $\rho$. The size of the computational domain is $L_x = 4 \pi h$ and $L_z = 2 \pi h$ in the streamwise and spanwise directions, discretised by $N_x = N_z = 256$, $512$ and $1024$ Fourier modes (further increased by a factor 3/2 for de-aliasing). In the wall-normal direction the differential operators are discretised via fourth-order compact finite differences using respectively $ N_y = 256 $, $250$ and $500$ points collocated on a non-uniform grid. Further details are provided in table~\ref{tab:discretization-parameters}. In this table and throughout the whole paper, quantities denoted with the superscript $+$ are given in viscous units, i.e. normalised with $u_\tau$ and $\nu$.

The database for the flow around around a finite rectangular cylinder is taken from the DNS study by \cite{cimarelli-leonforte-angeli-2018}, where the information on the numerical setup can be found. 
A rectangular cylinder of length $5h$, thickness $h$ and indefinite span is immersed in a uniform flow with free-stream velocity $U_\infty$ aligned with the $x$ direction. The Reynolds number is $Re = U_\infty h / \nu = 3000$. The streamwise, wall-normal and spanwise size of the computation domain is $\left(L_x, L_y, L_z \right) = \left( 112h, 50h, 5h\right)$. The leading edge of the cylinder is located $35h$ past the inlet of the computational box. The fluid domain is discretised through a Cartesian grid consisting of $1.5 \cdot 10^7$ hexahedral cells. The average resolution in the three spatial direction is $\left(\Delta x^+, \Delta y^+, \Delta z^+ \right) = \left(6.1, 0.31, 5.41 \right)$.

The AGKE terms are computed with an efficient code specifically developed for the present work, which extends a recently written code for the computation of the GKE equation \citep{gatti-etal-2019}. The symmetries described in Appendix \ref{sec:symmetries} are exploited to minimise the amount of memory required during the calculations. Each term of equations \eqref{eq:phi_r_gen}, \eqref{eq:phi_c_gen} and \eqref{eq:xi_gen} is decomposed into simpler correlation terms, which are then computed as products in Fourier space along the homogeneous directions, with huge savings in computing time. For maximum accuracy, derivatives in the homogeneous directions are computed in the Fourier space, otherwise a finite-differences scheme with a five-points computational stencil is used. Finally, a parallel strategy is  implemented \citep[see][for details]{gatti-etal-2019}. The calculation receives in input the fluctuating velocity field for each snapshot of the databases. It outputs $ \aver{\delta u_i \delta u_j}$, the flux vectors ${\bm{\phiC}}_{ij}$ and ${\bm{\phiR}}_{ij}$, and the various contributions to the source term $\src[ij]$ as in equation \eqref{eq:xi_gen} for each of the six different second-order structure functions, and in the whole physical and scale space. 

The statistical convergence of the data is verified by ensuring that the residual of equation \eqref{eq:AGKE_gen} is negligible compared to the dissipation, production and pressure-strain terms. 

\section{Example: the near-wall turbulence cycle}
\label{sec:inner}

A turbulent channel flow at $Re_\tau=200$ is considered in the following. The mean velocity vector is $\bm{U}\left(y \right) = \left\{ U(y), 0, 0\right\}$, directed along the streamwise direction $x = x_1$ and varying only with the wall-normal coordinate $y = x_2$; $z=x_3$ is the spanwise direction, and $u=u_1$, $v=u_2$ and $w=u_3$ indicate the three fluctuating velocity components. Since $y$ is the only direction of statistical inhomogeneity, $\aver{\duiduj}\left(Y, \bm{\R} \right)$ and all AGKE terms are function of the physical space only through the spatial coordinate $Y = \left(y + y^\prime \right)/2$, while still depending upon the whole scale vector $\bm{\R}$. Similarly, spatial transport of $\aver{\duiduj}$ occurs along $Y$ through the only nonzero component of the spatial flux $\phiC_{ij} = \phiC_{Y,ij}$.

The GKE for the scale energy $\aver{\del_i^2}$ has been thoroughly discussed in literature, \citep[see e.g.][]{marati-casciola-piva-2004,cimarelli-deangelis-casciola-2013,cimarelli-etal-2015,cimarelli-etal-2016}, and different interpretations and visualisation techniques have been suggested. For this reason, in the following we only address the new information offered by the AGKE. This includes the analysis of the anisotropic scale-energy redistribution operated by the pressure-strain terms, and that of the budget equation for $\aver{-\delta u \delta v}$. 
The analysis is also restricted to the subspace $r_x=0$: this is motivated by the turbulent vortical structures in channel flow being predominantly aligned in the streamwise direction. Such structures typically induce the largest negative correlation of velocity components for $r_x=0$ and characteristic values of $r_z$. A classic example are the so-called near-wall streaks, for which $r_z^+ \approx 60$. As a consequence of \eqref{eq:sf-corr}, the local maxima of, for instance, $\aver{\del\del}$ and terms appearing in its budget equation also occur for $r_x \approx 0$. Note that in the $r_x=0$ space the terms of the AGKE are not defined below the $Y=r_y/2$ plane, owing to the finite size of the channel in the wall-normal direction.

\subsection{Scale-energy redistribution by pressure strain}
\label{sec:innerstrain}

\begin{table}
\begin{tabular*}{1.0\textwidth}{@{\extracolsep{\fill}} l c c c c c c c c}
        &  \multicolumn{2}{c}{$\aver{\del_i\del_j}^+_m$} & \multicolumn{2}{c}{$\xi_{ij,m}^+$} & \multicolumn{2}{c}{$|\Pi_{ij}^+|_m$} & \multicolumn{2}{c}{$P_{ij,m}^+$} \\
        &  value             & position                  & value            & position        & value        & position              & value        & position                 \\[10pt]
$i=j=1$ & $17.15$            & $(0,58,14)$               & $0.74$           & $(0,39,12)$     & $0.14$       & $(0,50,24)$  & $1.24$ & $(0,39,12)$    \\
$i=j=2$ & $1.76$             & $(0,59,53)$               & $0.038$          & $(26,0,36)$     & $0.068$      & $(35,0,40)$  & $-$    & $-$             \\
$i=j=3$ & $2.83$             & $(42,0,45)$               & $0.053$          & $(0,42,9)$      & $0.12$       & $(0,46,12)$  & $-$    & $-$            \\
\end{tabular*}
\caption{Maximum values for diagonal terms of $\aver{\del_i\del_j}^+$, its source $\xi_{ij}^+$, absolute pressure strain $|\Pi_{ij}^+|$ and production $P_{ij}^+$ and positions in the $(\R_y^+,\R_z^+,Y^+)$-space.}
\label{tab:duidui}
\end{table}

The pressure-strain term $\Pi_{ij}$ redistributes energy among the diagonal components of $\aver{\del_i \del_j}$. Hence, at different scales and positions this term can be a source or a sink depending on its sign. To better understand its behaviour and link it to physical processes, it is instructive to briefly analyse the scales and position at which $\aver{\delta u \delta u}$, $\aver{\delta v \delta v}$, $\aver{\delta w \delta w}$ and their sources $\xi_{ij}$ are important.

The position and the intensity of the maxima, hereinafter denoted with the subscript $m$, of the diagonal components of $\aver{\del_i \del_j}$ and of the associated $\xi_{ij}$ and $\Pi_{ij}$ are reported in table \ref{tab:duidui}. $\aver{\delta u \delta u}$, $\aver{\delta v \delta v}$ and $\aver{\delta w \delta w}$ peak at small scales within the buffer layer, similarly to $\aver{\delta u_i^2}$ \citep{cimarelli-etal-2016}, with $\aver{\delta v \delta v}_m$ located further from the wall. The anisotropy of the flow is denoted, for instance, by $\aver{\delta w \delta w}_m$ being much lower than $\aver{\delta u \delta u}_m$ and occurring at $r_z=0$ and small $r_y$, whereas the other maxima occur at $r_z \ne 0$ and $r_y =0$. This difference is explained by the quasi-streamwise vortices populating the near-wall cycle \citep{schoppa-hussain-2002}: they induce negatively correlated regions of spanwise fluctuations at $r_y \ne 0$ and of streamwise and wall-normal fluctuations at $r_z \ne 0$.

The region of negative source terms partially coincides with the one of the source term in the GKE \citep[see e.g.][]{cimarelli-etal-2016}. As in the GKE, negative sources are observed at the lower boundary $Y={\R_y}/2$, and in the whole channel height at $\R_y, \R_z \rightarrow 0$: viscous dissipation dominates near the wall and at the smallest scales. However, the regions of large positive sources vary significantly among the three diagonal components (see table \ref{tab:duidui}). This is due to the different nature of the positive source of the three diagonal components of $\aver{\del_i \del_j}$. Indeed, in a turbulent channel flow the streamwise fluctuations are fed by the energy draining from the mean flow (i.e. by the production term $P_{11}$), whereas the cross-stream fluctuations are produced by the redistribution processes (i.e. the pressure-strain term $\Pi_{22}$ and $\Pi_{33}$). This explains also the larger order of magnitude of $\xi_{11,m}$. Unlike the GKE, the scale and space properties of this energy redistribution can be extracted from the AGKE (see equation \ref{eq:xi_gen}).

\begin{figure}
\centering
\includegraphics[width=\textwidth]{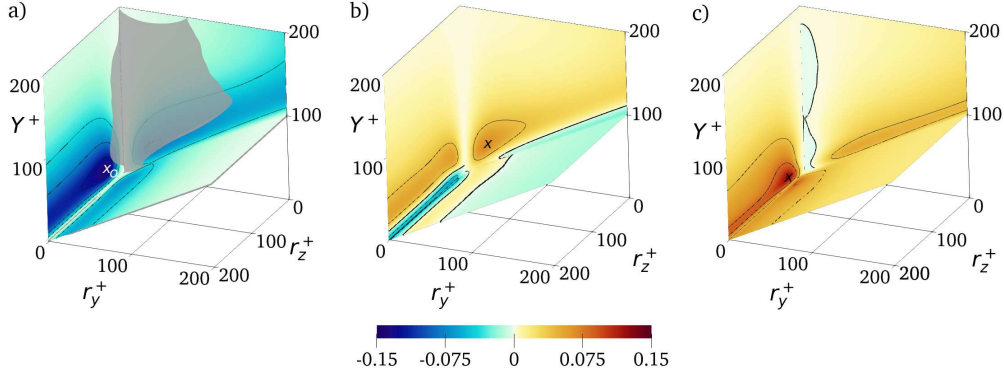}
\caption{Colour plot of: $\Pi_{11}^+$ (a), $\Pi_{22}^+$ (b) and $\Pi_{33}^+$ (c) on the bounding planes $r_y^+=0$, $r_z^+=0$ and $Y^+=r_y^+/2$. The contour lines increment is 0.04, with level zero indicated by a thick line. The two symbols identify the positions of the maxima of $\Pi_{ij}$ (cross) and $P_{ij}$ (circle). The isosurface in (a) corresponds to $\Pi_{22}/\Pi_{11}=-0.5$ (or equivalently $\Pi_{33}/\Pi_{11}=-0.5$), with $\Pi_{22}/\Pi_{11}<-0.5$ for smaller scales.}
\label{fig:duidui-Pstrain}
\end{figure}

Figure \ref{fig:duidui-Pstrain} plots the pressure-strain term for the diagonal components, with values and positions of their maxima as reported in table \ref{tab:duidui}. The figure shows the location of the pressure-strain maximum in absolute value together with the maximum production. Large values of $P_{11}$ occur near the plane $Y^+ = \R_y^+/2 +14$, except for the smallest scales in the region $\R_y^+<30$ and $\R_z^+<20$. 
On the other hand, $\Pi_{11}$ is negative almost everywhere, showing that the streamwise fluctuations lose energy at all scales to feed the other components. In particular, large negative values of $\Pi_{11}$, albeit much smaller than $P_{11}$, are seen near the plane $Y^+={\R_y^+}/2+24$, except for the region $\R_y^+, \R_z^+<20$. This brings to light the dominant scales and wall distances involved in the process of redistribution of $\aver{\delta u \delta u}$ towards the other components, and discriminates them from those involved in its production. On the contrary, at the smallest scales where viscous dissipation is dominant production and redistribution are not observed.

The pressure-strain terms of the cross-stream components, $\Pi_{22}$ and $\Pi_{33}$, are positive almost everywhere; they show a positive peak near the wall and remain larger than dissipation in different regions of the $r_x=0$ space. Their maxima are located in the vicinity of the plane $Y^+ = \R_y^+/2 +40$ for $\Pi_{22}$ and $Y^+ = \R_y^+/2 + 14$ for $\Pi_{33}$, where $\Pi_{11}$ is negative. Hence, at these scales and wall-normal distances $\aver{\delta u \delta u}$ loses energy towards $\aver{\delta v \delta v}$ and $\aver{\delta w \delta w}$. Moreover, $\Pi_{22}$ is negative in the very near-wall region, $Y^+< \R_y^+/2+5$, owing to the non-penetration wall boundary condition which converts $\aver{\del[v]\del[v]}$ into $\aver{\del[u]\del[u]}$ and $\aver{\del[w]\del[w]}$. Indeed, here $\Pi_{11}$ and $\Pi_{33}$ are positive. This phenomenon is known as the splatting effect \citep{mansour-kim-moin-1988}, and shows no scale dependency.

Different values of $\Pi_{22}$ and $\Pi_{33}$ imply an anisotropic redistribution of the streamwise fluctuations to the other components. Owing to the incompressibility constraint, the following relationship holds:
\begin{equation}
\frac{\Pi_{22}}{\Pi_{11}}+\frac{\Pi_{33}}{\Pi_{11}}=-1 \, .
\end{equation}
Hence, $\Pi_{22}/\Pi_{11} = \Pi_{33}/\Pi_{11} = -0.5$ corresponds to isotropic transfer of energy from the streamwise fluctuations towards the other components. In figure \ref{fig:duidui-Pstrain}a the isosurface $\Pi_{22}/ \Pi_{11}=-0.5$ is shown. The inner side at small scales of this surface is characterised by $\Pi_{22}/ \Pi_{11}<-0.5$, and thus by $\Pi_{22}>\Pi_{33}$ (as long as $\Pi_{11}<0$). Hence, at small scales the pressure strain preferentially redistributes streamwise energy to the vertical fluctuations. On the contrary, on the outer side of the surface $\Pi_{33}>\Pi_{22}$ holds, implying that at larger scales the streamwise energy is preferentially redistributed towards spanwise fluctuations. 

\subsection{Scale-by-scale budget of the off-diagonal term $\aver{-\delta u \delta v}$}

The only off-diagonal term associated with a nonzero component of the Reynolds stress tensor is $\aver{-\delta u \delta v}$ which, unlike the diagonal terms, is not definite in sign. Therefore, $\aver{-\delta u \delta v}$ and its fluxes cannot be interpreted in terms of energy and energy transfer. $\aver{-\delta u \delta v}$ describes the statistical dependence or, more precisely, the correlation between $\del[u]$ and $\del[v]$ and, for large $\bm{\R}$, the mean momentum transfer. Concepts as production and dissipation only apply to the source term $\src[12]$ after the sign of $\aver{-\delta u \delta v}$ is taken into account. 

\subsubsection{Intensity, production and redistribution}

\begin{table}
\begin{tabular*}{1.0\textwidth}{@{\extracolsep{\fill}} c c c c c c c c c c }
\multicolumn{2}{c}{$\aver{-\delta u \delta v}^+_{max}$} &\multicolumn{2}{c}{$\xi_{12,max}^+$} & \multicolumn{2}{c}{$\xi_{12,min}^+$} &\multicolumn{2}{c}{$\Pi_{12,min}^+$} & \multicolumn{2}{c}{$P_{12,max}^+$}  \\
value             & position             &     value            & position       & value        & position           &   value        & position  & value & position   \\[5pt]
$2.06$            & $(0,53,30)$          & $0.089$          & $(0,20,12)$        & $-0.093$     & $(19,0,12)$        &  $-0.183$     & $(0,30,17)$&$0.197$&$(22,0,22)$ \\
\end{tabular*}
\caption{Maximum value for $\aver{-\delta u \delta v}^+$, maximum and minimum for the source $\xi_{12}^+$, minimum for the pressure strain $\Pi_{12}^+$ and maximum of the production $P_{12}^+$ and positions in the $(\R_y^+,\R_z^+,Y^+)$-space.}
\label{tab:dudv}
\end{table}

\begin{figure}
\centering
\includegraphics[width=\textwidth]{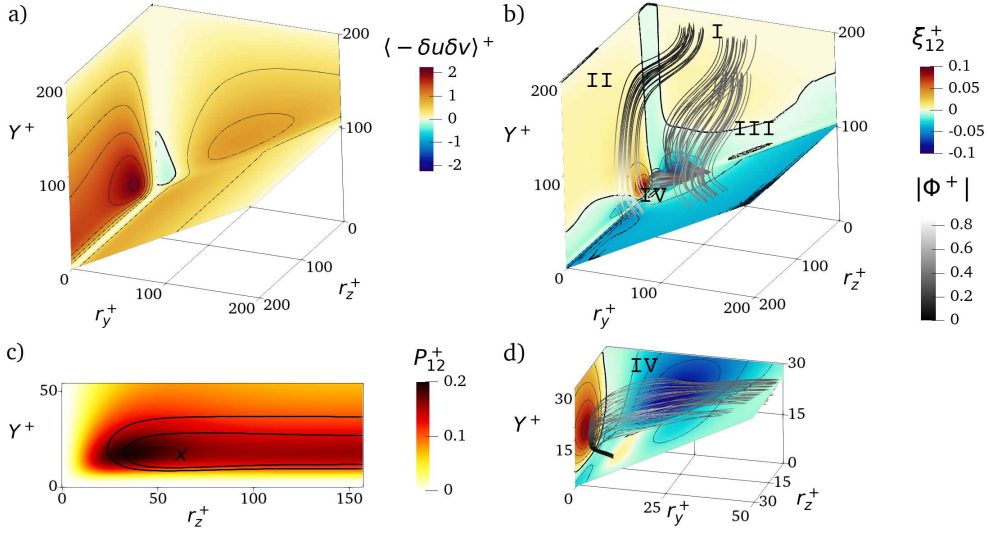}
\caption{Colour plot of $\aver{- \delta u \delta v}$ and its budget terms in the 3-dimensional space $r_x=0$. 
(a) $\aver{-\delta u \delta v}^+$: contour lines increment by 0.4, with zero indicated by a thick line.
(b) Colour plot of $\src[12]^+$: contour lines increment by 0.02, with zero indicated by a thick line. The gray lines are tangent to the flux vector $(\phiR_y,\phiR_z,\phiC)$ and coloured with its magnitude. A zoom of the region near the origin is shown in panel (d).
(c) Colour plot of $P_{12}^+$ in the $\R_x=\R_y=0$ plane, with isolines for $\Pi_{12}^+$ demonstrating the different scales involved and the different position of the maximum. The X symbol locates the position of the maximum for $\Pi_{12}$.
}
\label{fig:dudv}
\end{figure}

The off-diagonal term $\aver{-\delta u \delta v}$ and its budget are plotted in figure \ref{fig:dudv}, and corresponding quantitative information is reported in table \ref{tab:dudv}. As shown by figure \ref{fig:dudv}a, $\aver{-\delta u \delta v}$ is positive almost throughout the entire physical/scale space except at very small separations ($\R_z^+ =0, \R_y \leq 10$) for $Y^+<50$. The largest positive values of $\aver{-\delta u \delta v}$ are in the buffer layer at $15 \leq Y^+ \leq 60 $, at spanwise scales $40 \leq \R_z^+ \leq 80$ and vanishing $r_y$. A second, less prominent local maximum of $\aver{-\delta u \delta v}$ is located near the $r_z=0$ plane. 

The source term $\src[12]$, plotted in figure \ref{fig:dudv}b, is dominated by the (positive) production term $P_{12}$ and the (negative) pressure-strain term $\Pi_{12}$ (see equation \eqref{eq:AGKE_uv} for their definitions). Indeed, the viscous pseudo-dissipation $D_{12}$ plays a minor role, as in the single-point budget for $\aver{-uv}$ \citep[see e.g.][]{mansour-kim-moin-1988}. Large positive and negative values of $\src[12]$ define two distinct regions in the buffer layer (figure \ref{fig:dudv}d). The positive peak corresponds to spanwise scales $10 \le \R_z^+ \le 50$, while the negative one to small scales ($\R_z^+ \approx 0$). Moreover, $\src[12]$ is negative in a portion of the $Y^+ = \R_y^+/2$ plane, implying that turbulent structures extending down to the wall are inactive in the production of $\aver{-\delta u \delta v}$.

\begin{figure}
\includegraphics[]{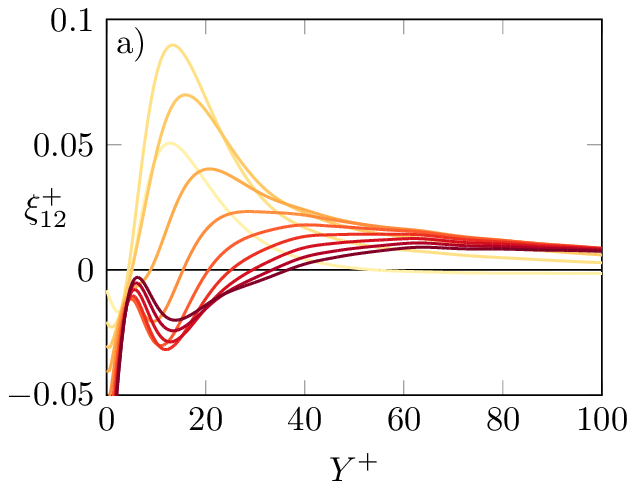}
\includegraphics[]{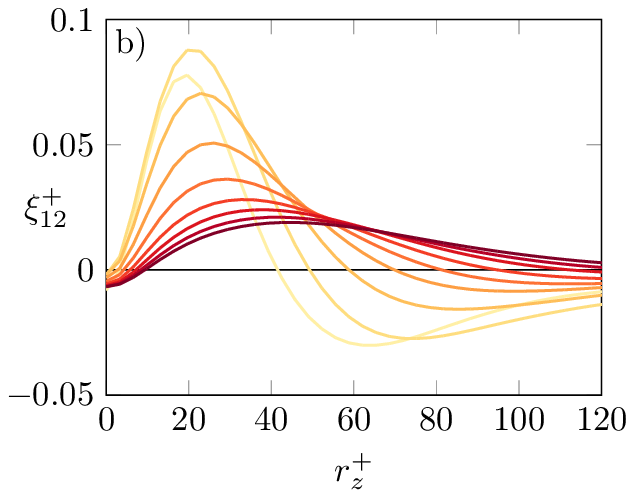}
\caption{Source term $\xi_{12}$ in the $\R_y=0$ plane. (a) $\xi_{12}^+$ versus $Y^+$ for different $\R_z^+=\left(10:10:100\right)$. (b) $\xi_{12}^+$ versus $\R_z^+$ for different $Y^+=\left(10:5:50\right)$. Line colours encode the value of the parameter, which increases from yellow (light) to red (dark).} 
\label{fig:xi12-ry0-rx0}
\end{figure}

It is worth noting that $\src[12]$ strongly varies with spanwise separation, as seen in the $r_y=0$ plane (figure \ref{fig:dudv}c; see also figure \ref{fig:xi12-ry0-rx0}). In comparison to the global picture obtained from single-point analysis of $\aver{-uv}$ in the buffer layer (here recovered in the limit $r_z \rightarrow L_z/2$) where the source term is slightly negative, one can additionally appreciate the existence of a large positive peak of $\src[12]$ at $r_z^+=20$ and a negative one at $r_z^+=70$ (figure \ref{fig:xi12-ry0-rx0}b). Indeed, $P_{12}$ and $\Pi_{12}$ are of the same order of magnitude throughout the $r_y=0$ plane, but reach their extreme values at different spanwise scales, see figure \ref{fig:dudv}c. In particular large values of $P_{12}$ are found at $(\R_z^+,Y^+) \approx (30,17)$, whereas large negative values of $\Pi_{12}$ are found at $(\R_z^+,Y^+) \approx (60,16)$. The structural interpretation of these findings is discussed below in \S\ref{sec:dudvqsv}. 

\subsubsection{Fluxes}

The transfer of $\aver{-\delta u \delta v}$ in space and among scales is determined by the flux vector $(\phiR_y,\phiR_z,\phiC)$, and is visualised via its field lines. These field lines can be grouped in two families. The lines of the first family enter the domain from the channel centerline, $Y=h$, and descend towards the wall; they can be further grouped in sets I, II and III as shown in figure \ref{fig:dudv}b. The second family only contains set IV, and is visible in the zoomed figure \ref{fig:dudv}d; its field lines are confined to the near-wall region, and connect the positive and negative peaks of $\xi_{12}$.
\begin{figure}
\centering
\includegraphics[]{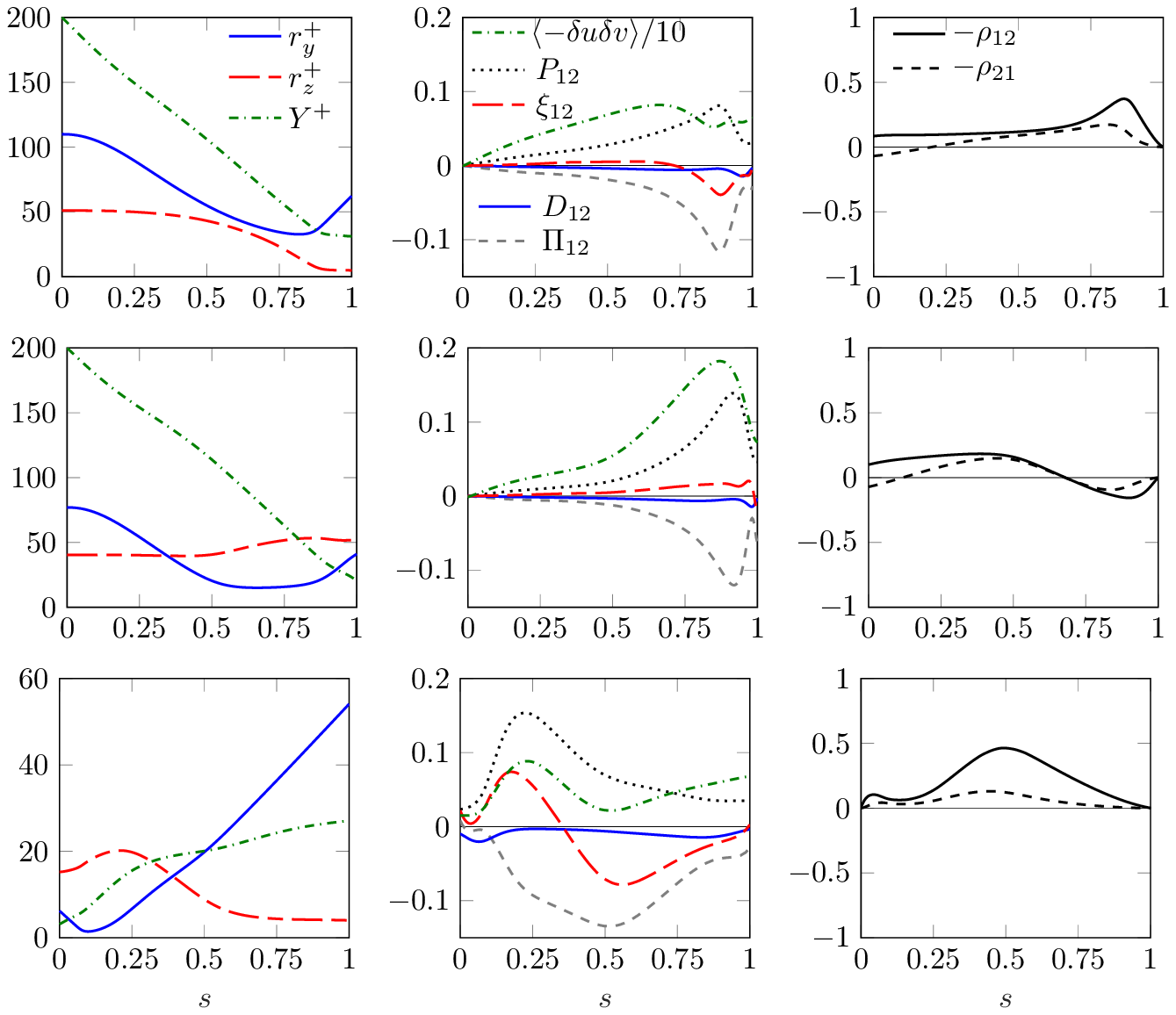}
\caption{Field lines of the flux vector for $\aver{-\delta u \delta v}$. Top: set I; centre: set II; bottom: set IV. Left: evolution of the values of $Y$ (\protect\linesample{black!50!green}{dashdotted}), $\R_y$ (\protect\linesample{blue}{solid}), $\R_z$ (\protect\linesample{red}{dash pattern=on 10pt off 3pt on 5pt off 3pt}),
 along a representative field line as a function of its dimensionless arc length $s$. Centre: values of $\aver{-\delta u \delta v}/10$ (\protect\linesample{black!50!green}{dashdotted}), $\src[12]$ (\protect\linesample{red}{dash pattern=on 10pt off 3pt on 5pt off 3pt}), $P_{12}$ (\protect\linesample{black}{dotted}), $D_{12}$ (\protect\linesample{blue}{solid}) and $\Pi_{12}$ (\protect\linesample{gray}{dashed}) along the line. Right: Evolution of $-\rho_{12}$ (\protect\linesample{black}{solid}) and $-\rho_{21}$ (\protect\linesample{black}{dashed}) along the line.}
\label{fig:dudv_field-lines}
\end{figure}

Various quantities can be tracked along representative field lines, as done in figure \ref{fig:dudv_field-lines}. The position along a field line of length $\ell$ in the $(r_z, r_y, Y)$ space is described by the normalised curvilinear coordinate
\begin{equation}
s=\frac{1}{\ell} \int_{0}^{\ell} \mathrm{d}s \qquad \mbox{with} \qquad \mathrm{d}s=\sqrt{d \R_z^2+ d \R_y^2+ d Y^2} \,.
\label{eq:s-definition}
\end{equation}
The values of $\R_y$, $\R_z$ and $Y$ 
(see figure \ref{fig:sf-sketch}) are plotted in the left column of the figure; the central column plots the evolution of $\aver{-\delta u \delta v}$, $\src[12]$, $P_{12}$, $\Pi_{12}$ and the pseudo-dissipation $D_{12}$ along the line; the right column plots the evolution of the correlation coefficient $\rho_{ij}$ defined by 
\begin{equation}
\rho_{ij}=\frac{R_{ij}(Y,\R_y,\R_z)}{\sqrt{\aver{u_i u_i}(Y) \aver{u_j u_j}(Y)}}
\label{eq:rho}
\end{equation}
where repeated indices do not imply summation. $R_{ij}$ is linked to $\aver{\duiduj}$ by equation \eqref{eq:sf-corr}.

The top and central panels of figure \ref{fig:dudv_field-lines} illustrate the evolution of various quantities along representative lines of set I and II. Both lines are qualitatively similar: they highlight a transfer of $\aver{-\delta u \delta v}$ from the centerline to the near-wall region, through first decreasing and then increasing wall-normal scales. At the centreline they are parallel to the $Y$ axis, consistently with the AGKE symmetries (see appendix \ref{sec:symmetries}). However, lines of set I are attracted by the negative peak of $\src[12]$ towards smaller $\R_z$, while those of set II are repulsed from the positive source peak towards larger $\R_z$. Lines of set III are not shown for the sake of brevity, since they pass through regions of large separations and are characterised by almost zero correlation, see equation (\ref{eq:rho}). On the other hand, lines of sets I and II exist at smaller $\R_y$ and $\R_z$ and, as shown in the upper-right and central-right panels of figure \ref{fig:dudv_field-lines}, are characterised by finite levels of correlation.
Along lines of set I and II, $\aver{-\delta u \delta v}$ increases from zero at the centerline (due to the AGKE symmetries) to reach a positive peak in the near-wall region. Similarly, $\src[12]$ shows a negative/positive peak when the lines of set I/II approach the near-wall region as the pressure-strain/production overcomes the production/pressure-strain.

The evolution of the correlation coefficients $-\rho_{12}$ and $-\rho_{21}$ (recall that $\rho_{ij} \ne \rho_{ji}$ for $i \ne j$, see equation \eqref{eq:rho}) is used to extract information about the turbulent structures involved in production, transfer and dissipation processes highlighted along the lines. As shown in the left-top and left-central panels of figure \ref{fig:dudv_field-lines}, at values of the curvilinear coordinate $s>0.75$ corresponding to $Y^+<60$, lines of set I intersect positive $-\rho_{12}$ and $-\rho_{21}$ for small $\R_z$ and $\R_y$, while those of set II intersect negative correlations at larger $\R_z^+ \sim 50$ and smaller $\R_y$. For both sets, this is consistent with the flow field induced by near-wall quasi-streamwise vortices, creating positive and negative cross-correlation at values of separation in agreement with the present analysis; positive $-\rho_{12}$ is associated to $u$ and $v$ fluctuations at the same-side of the vortices (i.e. small $\R_z$), whereas negative $-\rho_{12}$ is associated to opposite-side fluctuations. Hence, we relate the peaks of $P_{12}$ and $\Pi_{12}$ (and consequently of $\src[12]$) along the lines of set I and II to such structures.

The lines of set IV, shown in figures \ref{fig:dudv}b and \ref{fig:dudv}d and in the bottom panels of figure \ref{fig:dudv_field-lines}, behave differently. The field lines originate in the lower boundary of the domain at $(\R_y^+,\R_z^+,Y^+)=(6,15,3)$. Along their path they first intercept the positive peak of $\src[12]$ at small $\R_y$ where $\aver{- \delta u \delta v}$ is maximum. Then, they pass through the negative peak of $\src[12]$, located at smaller $r_z$ and larger $r_y$, where $\aver{-\delta u \delta v}$ is smaller. Eventually, they again vanish in the lower boundary of the domain.

Focusing on the correlation coefficient $-\rho_{12}$, lines of set IV intersect a positive value along their complete extension. In detail, the lines first intersect small values of $-\rho_{12}$ for $r_z^+ \approx 20$ and $Y^+<5$ and then larger $-\rho_{12}$ for smaller $r_z^+$ and larger $Y^+$. Hence, this set of lines highlights a transfer of $\aver{-\delta u \delta v}$ between the small $uv$-structures created in the viscous sublayer by the wall boundary condition \citep{sillero-jimenez-moser-2014} and the turbulent structures of the near-wall cycle.

\subsubsection{Structural properties of wall turbulence}
\label{sec:dudvqsv}

\begin{figure}
\centering
\includegraphics[]{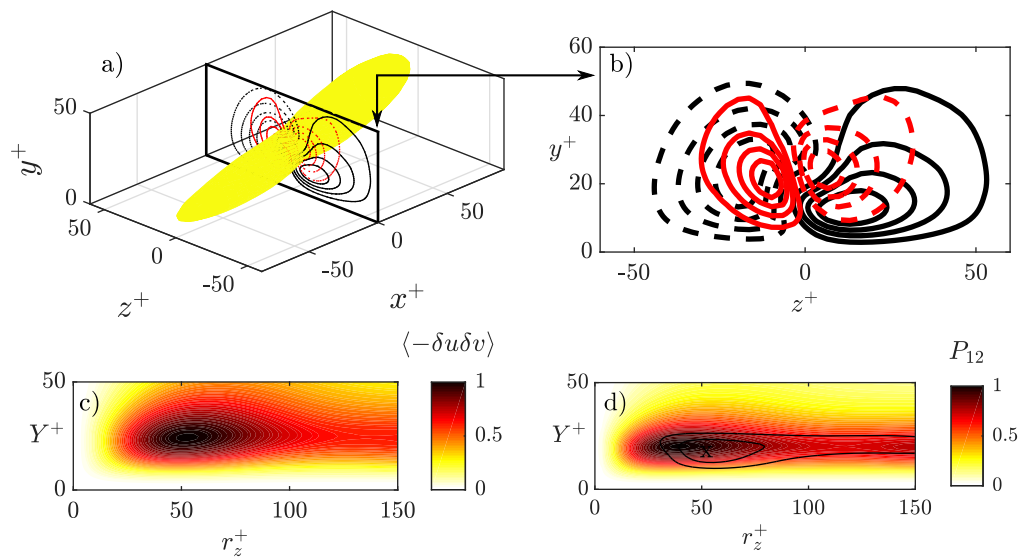}
\caption{(a) Ensemble-average quasi-streamwise vortex, educed as described in \S\ref{sec:dudvqsv} and Appendix C, represented as isosurface $\lambda^+_{\mathrm{ci}}=0.145$ of the swirling strength criterion \citep{zhou-etal-1999}. The coherent streamwise (black) and wall-normal (red) velocity field induced by the vortex are plotted on a $y-z$ plane, located at $x=0$, passing through the centre of the vortex. 
The plane is represented both in (a) and, more in detail, in (b). Contour levels at (0.2:0.2:0.8) of the maximum (solid line) and of the minimum (dashed line) of the respective component ($0.0058$ and $-0.0077$ for $u$ and $0.0035$ and $-0.0035$ for $v$) are plotted on a $y-z$ plane passing through the centre of the vortex, located at $z=0$. (c) Colour map of the corresponding $\aver{\del[u]\del[v]}$ normalised by its maximum value on the plane $\R_x=\R_y=0$. (d) Colour map of the corresponding $P_{12}$, and contours of $\Pi_{12}$, normalised by its  maximum value, on the plane $\R_x=\R_y=0$. Contours levels are shown at ($-0.6$, $-0.7$), and the X symbols locates the maximum.}
\label{fig:uv-sketch}
\end{figure}

To connect the main statistical features of $\aver{-\delta u \delta v}$ in the buffer layer to the turbulent structures that populate it, we compute the $\aver{-\delta u \delta v}$ AGKE budget from the velocity field induced by the ensemble-averaged quasi-streamwise vortex. Such vortex, visualised in figure \ref{fig:uv-sketch}a, represents the characteristic near-wall coherent structure in the average sense. The procedure which extracts the ensemble-average vortical structure from the DNS database is very similar to the one presented by \cite{jeong-etal-1997}, which is slightly modified here to focus on the structures in the buffer layer only. Details of the procedure are provided in Appendix C.

The ensemble-averaged velocity field is shown in figure \ref{fig:uv-sketch}b in a $z^+-y^+$ plane passing through the vortex centre. The corresponding $\aver{-\del[u]\del[v]}$, normalised by its maximum in the $r_x=0$ space, is shown in figure \ref{fig:uv-sketch}c in the $r_x=r_y=0$ plane. $\aver{-\del[u]\del[v]}$ computed for the average structure shows a remarkable agreement with the same quantity computed for the turbulent channel flow. In particular, its maximum occurs at $\left(\R_y^+,\R_z^+,Y^+ \right)=\left(0, 52, 25 \right)$, i.e. nearly the same location $\left(\R_y^+,\R_z^+,Y^+ \right)=\left(0, 53, 30 \right)$ observed for the full velocity field (see table \ref{tab:dudv}). Figure \ref{fig:uv-sketch}d shows the production $P_{12}$ and the pressure-strain $\Pi_{12}$ normalised with the maximum production in the $r_x=0$ space. Again, the average quasi-streamwise vortex represents well the typical $r_z$ scales of production and pressure-strain of $\aver{-\del[u]\del[v]}$. The peak of $P_{12}$ occurs at $\left( \R_z^+, Y^+ \right)=\left(39.2, 20.0 \right)$ while the minimum of $\Pi_{12}$ is located at $\left( \R_z^+, Y^+ \right)=\left(52.3, 19.0 \right)$, i.e. at a larger spanwise scale, similar to what figure \ref{fig:dudv}c shows for the full velocity field.

\section{Example: the outer turbulence cycle}
\label{sec:outer}
\begin{figure}
\centering
\includegraphics[]{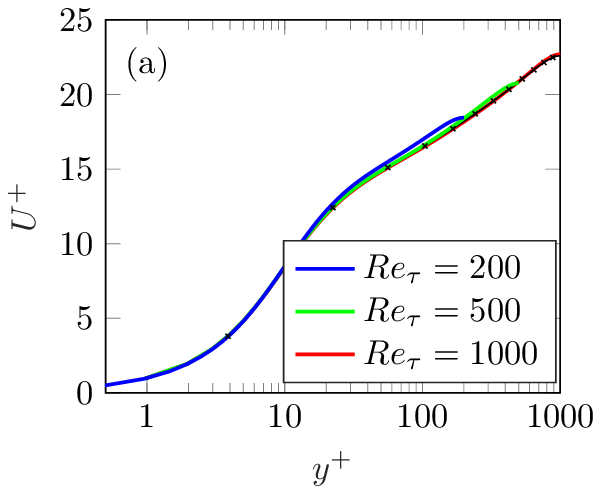}\hspace{10pt}\includegraphics[]{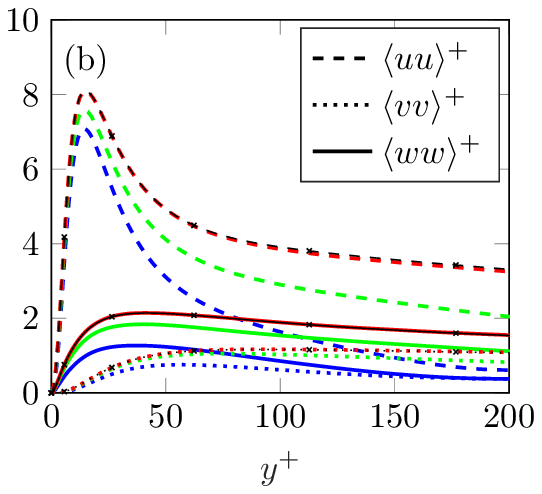}
\caption{Profile of (a) mean velocity and (b) velocity fluctuation variance  at $Re_\tau=200$, $500$ and $1000$. For validation, in both panels data from \cite{lee-moser-2015} at $Re_\tau=1000$ are also plotted with a thin black line connecting open symbols, which is nearly perfectly overlapping.}  
\label{fig:valid}
\end{figure}
Thanks to its ability to account for scales also in directions of statistical inhomogeneity, the AGKE becomes increasingly informative as the range of turbulent scales widens. For the turbulent channel flow, $Re_\tau$ is the ratio between the outer geometrical lengthscale $h$ and the inner viscous lengthscale $\nu/u_\tau$. Hence, for increasing values of $Re_\tau$, the typical scales of the autonomous near-wall cycle discussed in \S\ref{sec:inner} are constant in viscous units but shrink compared to $h$. Meanwhile, a whole new hierarchy of motions starts to appear: they include larger scales in the logarithmic region and form the so-called outer cycle \citep[see, for instance,][]{cossu-hwang-2017}. The wall-normal extent of such motions is typically not accounted for by other frameworks for the analysis of scale transfers, but can be easily studied by the AGKE. 

A comparative AGKE analysis for a channel flow at the three different values of $Re_\tau=200$, $500$ and $1000$ is presented below. The main features of the DNS databases have been already introduced in \S\ref{sec:DNS}. The profiles of mean velocity and variance of velocity fluctuations at all values of $Re$ considered in the following are reported in figure~\ref{fig:valid}, which confirms the full agreement of such statistics with the database available from \cite{lee-moser-2015}. 

\begin{figure}
\centering
\includegraphics[]{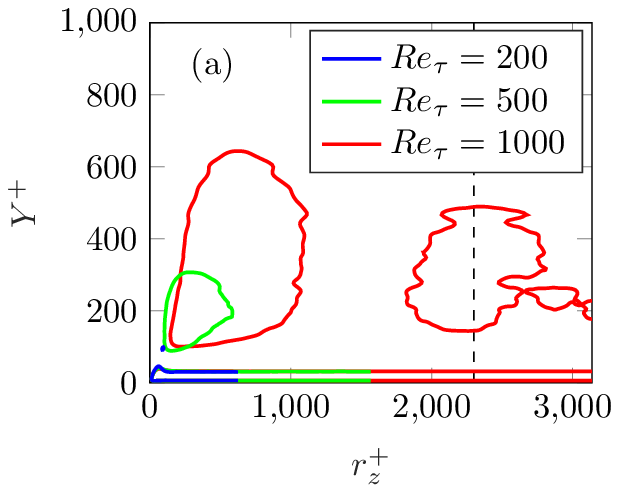}\includegraphics[]{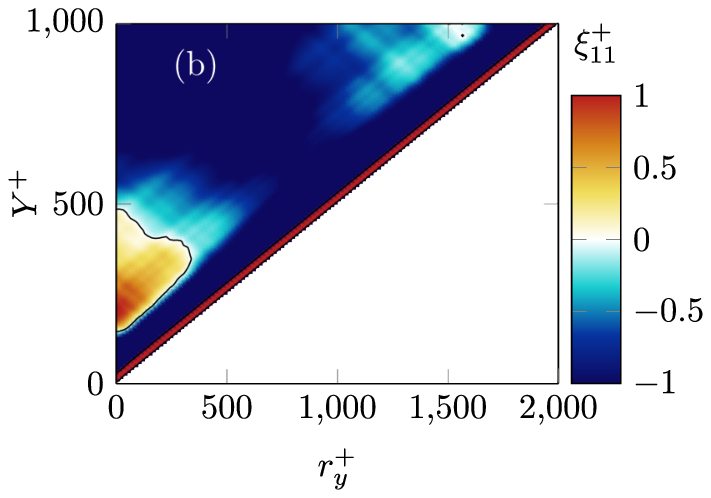}
\caption{(a) Contour $\xi_{11}=0$ for $Re_\tau=200$, $500$ and $1000$ in the $\left(r_z, Y \right)$ plane at $r_x=r_y=0$. (b) Color map of $\xi_{11}$ at $Re_\tau=1000$ in the $\left(r_y, Y \right)$ plane for $r_x=0$ and $r_z^+=2300$, i.e. for the plane shown in the left panel (vertical dashed line) which crosses the large-scale $\xi_{11}$ maximum at $Re_\tau=1000$.  \label{fig:re}}
\end{figure}
Figure~\ref{fig:re}a shows the contour $\src[11]=0$ in the $(r_z, Y)$ plane at $r_x=r_y=0$. Taking $r_y=0$ is equivalent to the classic approach, where only wall-parallel separations (or wavenumbers in the spectral analysis) are considered. Three different regions of net energy source $\src[11]>0$, enclosed by the isoline $\src[11]=0$, can be distinguished. The first region, which collapses for all values of $Re$ with viscous scaling, corresponds to the net production of $\aver{\del \del}$ within the near-wall cycle, already described in \S\ref{sec:inner}, and takes place at all spanwise separations. The second region of $\src[11] > 0$ is found for approximately $r_z^+ \leq h^+$ and $Y^+ \leq 0.6 h^+$. Here the left boundary of the contour $\src[11]=0$ represents the cross-over value of $r_z$, for a given $Y$, separating larger production scales from smaller inertial scales. The cross-over scale increases approximately linearly with the wall distance, in agreement with the overlap layer predictions of the attached-eddy model \citep{townsend-1976}. \cite{cimarelli-etal-2015} carry out a detailed analysis of the scaling properties of this second source region, albeit in terms of $\aver{\delta u_i \delta u_i}$, while \cite{marusic-monty-2019} discuss the attached-eddy model and its implications. This second region of $\src[11] > 0$ is observed also with the analysis based on one-dimensional premultiplied spectral budgets \citep[see, for instance, figure 5 in][]{lee-moser-2019}, although here it is clearly separated from the one of the near-wall cycle. It is also interesting to note that this region, albeit weak and confined to a tiny range of spanwise scales and wall-normal positions, is already apparent at $Re_\tau=200$, something that can not be observed as easily from one-dimensional spectra.

Only for the largest value $Re_\tau=1000$ considered here, a third region of $\src[11] > 0$ appears, with spanwise scales $2h^+ \leq r_z^+ \leq 3h^+$ and values of $Y^+$ pertaining to the logarithmic layer. This third region is related to the production by additional large-scale turbulent features, whose statistical footprint cannot be predicted by using the attached-eddy hypothesis \citep{marusic-monty-2019}. These motions have been named superstructures \citep{hutchins-marusic-2007} when found in boundary layers and Large Scale Motions (LSM) or Very Large Scale Motions (VLSM) \citep{guala-hommema-adrian-2006} when observed in turbulent channels, pipes and plane Couette flows. Henceforth we will adopt the acronym LSM, disregarding the slight differences in the definition of the three terms given in literature. LSM are important for two main reasons. First, their relative contribution to the total turbulent kinetic energy and Reynolds shear stress rapidly increases with $Re_\tau$ \citep{ganapathisubramani-longmire-marusic-2003}, making LSM one of the main players in the outer cycle and thus an obvious target for flow control. Second, LSM  modulate the inner cycle \citep{mathis-hutchins-marusic-2009} and superpose to the near-wall turbulence \citep{hoyas-jimenez-2006}, thus causing the failure of exact viscous scaling for several statistical quantities, such as for example the wall-normal profiles of the streamwise and spanwise velocity fluctuations. 

Figure~\ref{fig:re}b focuses on the $Re_\tau=1000$ case, and illustrates how  the AGKE can naturally consider scales in the wall-normal inhomogeneous direction, something particularly useful to describe the volume-filling LSM. Contours of $\src[11]$ at $Re_\tau=1000$ are plotted in the $(r_y, Y)$ plane for $r_x=0$ and $r_z^+=2300$, i.e. the spanwise scale at which LSM have been observed in Figure~\ref{fig:re}a. The results reveal the wall-normal distribution of the net positive source, i.e. net production of $\aver{\del\del}$, occurring at the scales of the LSM throughout the channel. Positive $\src[11]$ is observed for $150 \leq Y^+ \leq 0.5 h^+$ at wall-normal scales in the range $0 \leq r_y^+ \leq 400$, while the bottom part of the contours runs parallel to the line $Y^+ = r_y^+/2 + 150$, indicating that the wall-normal scales related to LSM are self-similar, contrary to the spanwise ones. The wall-normal location and scale at which $\src[11]$ is active agrees remarkably well with the wall-normal extent of LSM measured by \cite{madhusudanan-illingworth-marusic-2019} utilising high-Re DNS data and  linearised Navier--Stokes equations subject to stochastic forcing. Interestingly, positive $\src[11]$ at the LSM spanwise scale occurs also for $r_y^+ \approx 1.7 h^+$ and $Y^+ \approx h^+$ (see figure~\ref{fig:re}b), indicating that $\aver{\del\del}$ is also produced at very large wall-normal scales at the centerline and thus that large-scale negative correlation of the streamwise velocity fluctuations is produced across the two channel halves.



\section{Example: separating and reattaching flows}
\label{sec:barc}

The separating and reattaching flow over a rectangular cylinder with length-to-height ratio of 5 is a popular benchmark for bluff-body aerodynamics \citep{bruno-2014}, known as BARC. It is considered here as an example of complex flow with two inhomogeneous directions and multiple separations and reattachments. Various flow structures are known to exist in different parts of the main recirculating bubble, and recently it has been suggested \citep{cimarelli-leonforte-angeli-2018} that streamwise- and spanwise-oriented vortices populate the attached and detached portion respectively of the reverse boundary layer.

\begin{figure}
\centering
\includegraphics[width=0.9\textwidth]{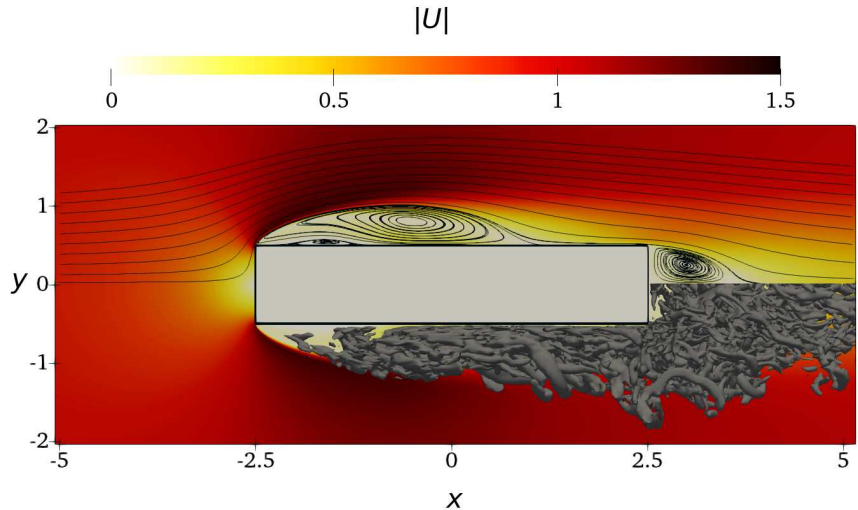}
\caption{Mean and instantaneous flow field around a 5:1 rectangular cylinder at $Re=3000$ (flow from left to right; $Re$ is based on free-stream velocity and cylinder height). The color background describes the mean velocity field $\bm{U}\left(x, y \right) = \left\{ U, V, 0\right\}$. In the upper half, mean streamlines show flow detachment at the sharp leading edge, a large recirculation bubble, a smaller secondary bubble and the rear separation in the wake. In the lower half, iso-surfaces for $\lambda_2 = -7$ visualize instantaneous vortical structures.}
\label{fig:lambda2}
\end{figure}
The snapshots used below for the AGKE analysis of the BARC flow are taken from the DNS study by \citet{cimarelli-leonforte-angeli-2018}. Figure~\ref{fig:lambda2} visualises the mean and instantaneous velocity fields. Three recirculation zones are present: a large-scale primary bubble originating from the leading-edge separation, a separation in the wake and a smaller secondary recirculation within the primary bubble. Separating and reattaching flows often feature the simultaneous presence of small scales, related to turbulent motions, and large scales, related to shedding of large-scale vortices. A full understanding of their interaction would be of paramount importance for the correct prediction and control of the flow \citep{kiya-1983, cherry-1984, kiya-1985, nakamura-1991, tafti-1991}. In particular, transition in the leading-edge shear layer is strongly affected by such multi-scale interactions: a region with negative turbulence production has been identified \citep{cimarelli-etal-2019-negative}, which leads to overwhelming difficulties with turbulence closures \citep{bruno-2014}. A key role is played by the turbulent structures advected within the main recirculating bubble, which trigger the transition of the leading-edge shear layer that in turn creates them, thus effectively belonging to a self-sustaining cycle. Remarkably, these structures appear to be quasi-streamwise vortices at the beginning of the reverse boundary layer and, while working their way upstream, become spanwise vortices. However, this process is far from being fully understood, and the AGKE will be used to clarify it. Note that, since statistical homogeneity only applies to the spanwise direction and time, all two-point statistics involved in the AGKE are now function of the separation vector $\bm{r}$, and the two spatial coordinates $X = \left( x + x^\prime \right)/2$ and  $Y = \left(y + y^\prime \right)/2$. In the figures that follow, lengths and velocities are made dimensionless with the free-stream velocity and the cylinder height.

\begin{figure}
\centering
\includegraphics[width=0.9\textwidth]{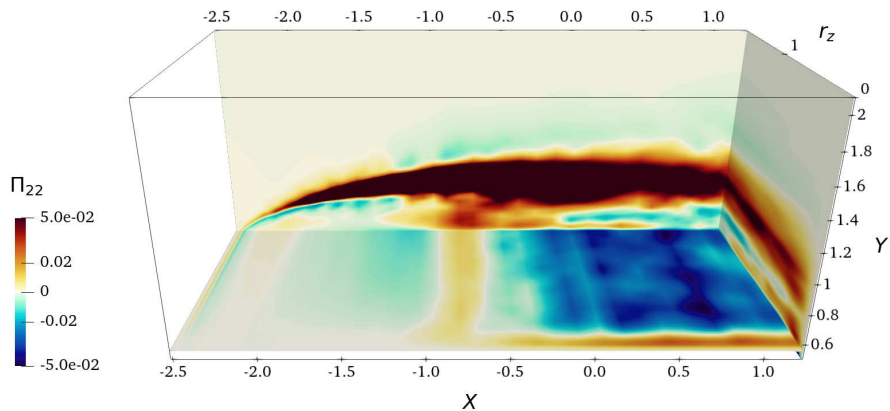}
\caption{Pressure-strain term $\Pi_{22}$ in the $\left(X,Y,r_z \right)$-space for $r_x=r_y=0$. Colour plot are shown on the planes $X=1.2$, $Y=0.56$ and $r_z=1.7$.}
\label{fig:barcPI22}
\end{figure}

We start with the component $\aver{\del[v]\del[v]}$, since it is the most obvious proxy for the local alignment of turbulent structures; in fact a streamwise structure would be revealed by a local maximum of $\aver{\del[v]\del[v]}$ at $r_x=0$ and a finite $r_z$, whereas a spanwise structure implies a local maximum at finite $r_x$ and $r_z=0$. In figure \ref{fig:barcPI22} the pressure-strain term $\Pi_{22}$ is shown in the $\left(X,Y,r_z \right)$ space that embraces the whole primary bubble for $r_x=r_y=0$. $\Pi_{22}$ is first observed to mark clearly the outer edge of the bubble. Within the bubble, $\Pi_{22}$ is highly scale- and position-dependent, and it differs from channel flow as discussed in \S\ref{sec:inner}. For instance, along the reverse attached boundary layer, i.e. for $-0.8 \le X \le 1$ and $Y \le 0.75$, $\Pi_{22}$ shows an evident positive peak at small spanwise scales ($r_z < 0.1$) even very near the wall, whereas in the channel flow the splatting effect leads to negative $\Pi_{22}$ (see figure~\ref{fig:duidui-Pstrain} in \S\ref{sec:inner}). Therefore, in this region $\Pi_{22}$ feeds clearly identified spanwise scales which are compatible with streamwise-aligned vortices. However, closer to the detachment of the reverse boundary layer (i.e. $-0.8 \leq X \leq -1.2$), an abrupt change takes place: $\Pi_{22}$ becomes positive at every spanwise separation, suggesting that once detached the reverse boundary layer is no longer populated by streamwise vortices. 

\begin{figure}
\centering
\includegraphics[width=0.495\textwidth,trim = 40mm 0mm 40mm 0mm, clip]{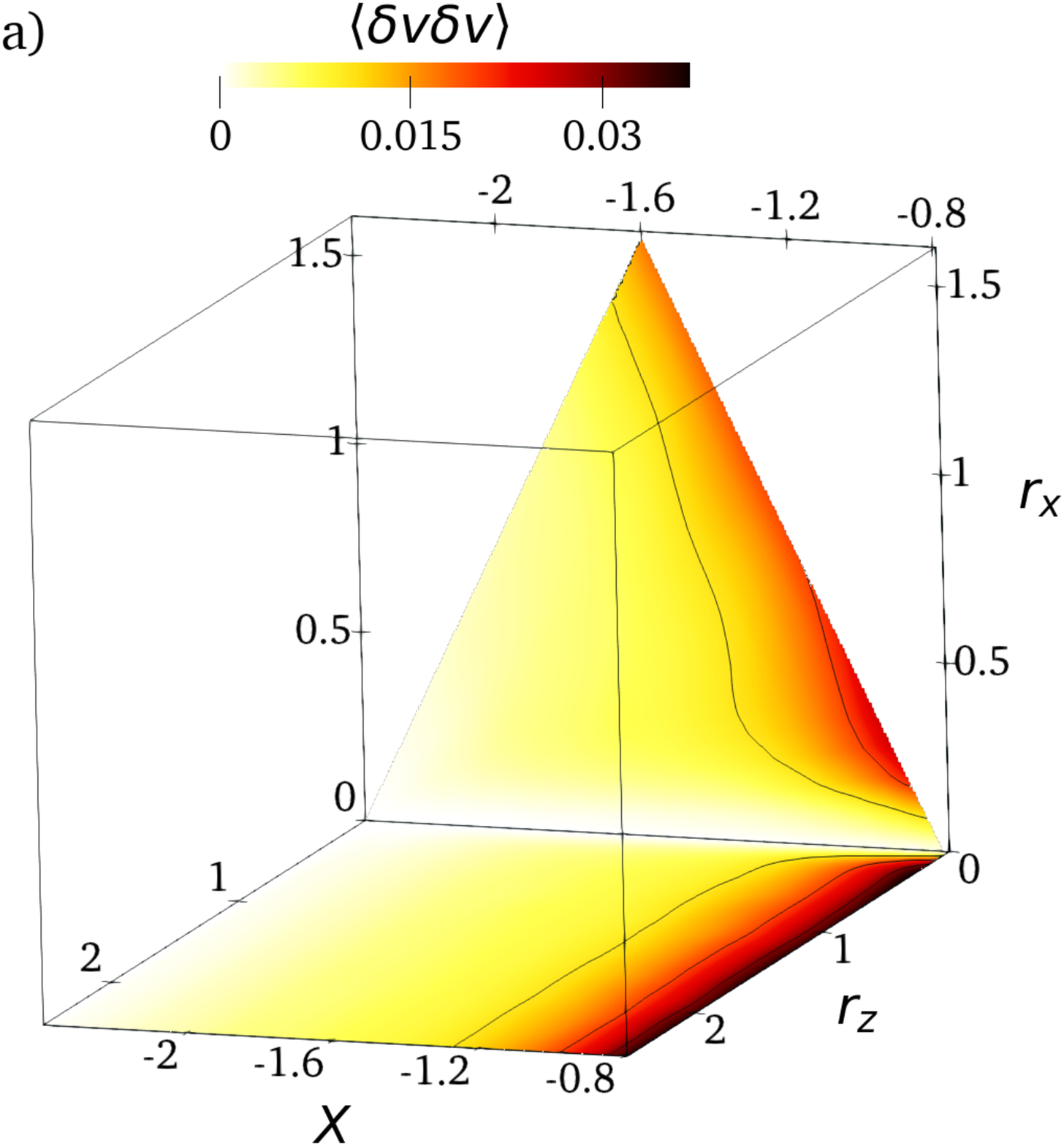}
\includegraphics[width=0.495\textwidth,trim = 40mm 0mm 40mm 0mm, clip]{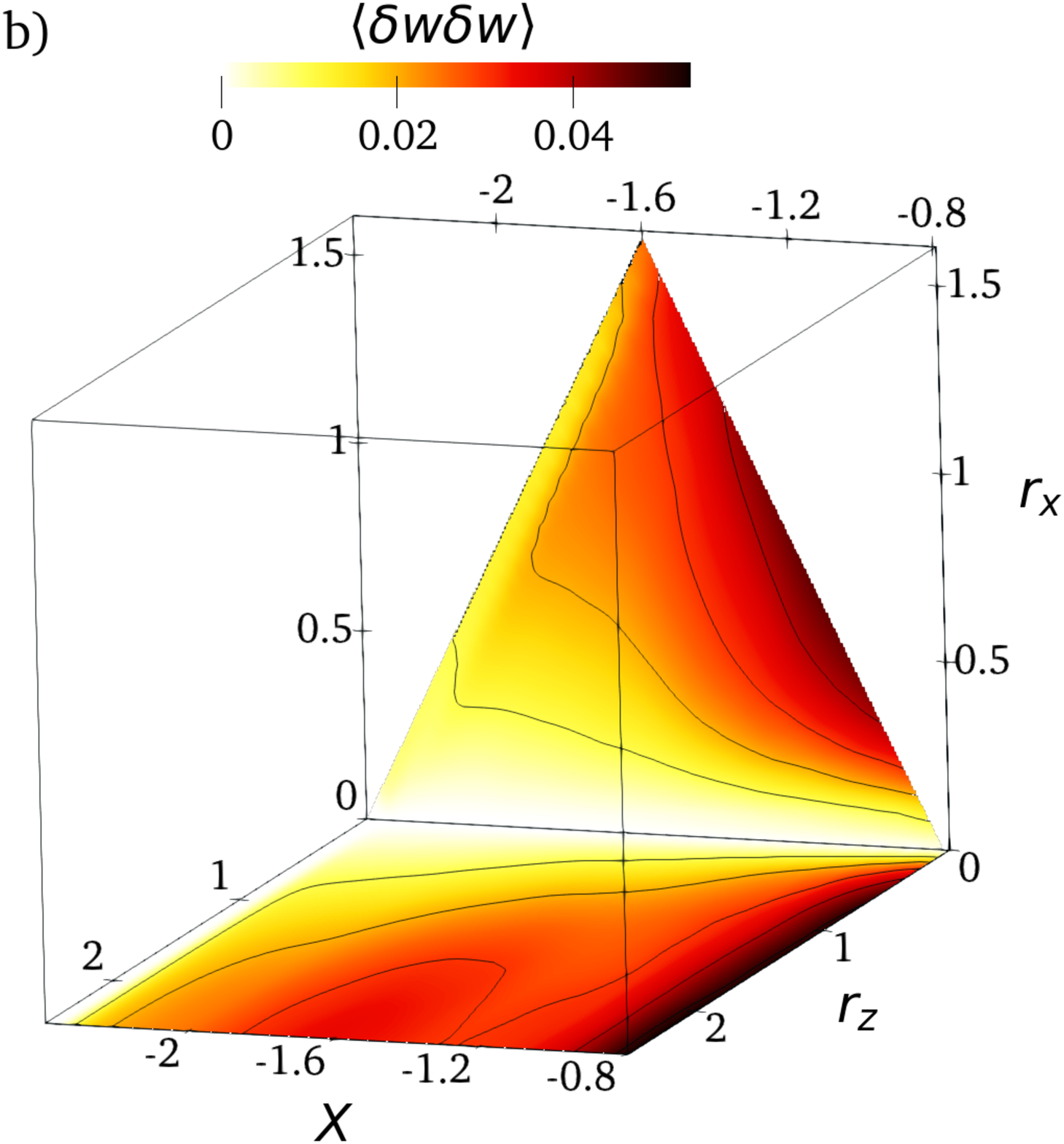}
\caption{$\aver{\delta v \delta v}$ (a) and $\aver{\delta w \delta w}$ (b) in the $(X,r_x,r_z)$-space for $r_y=0$ and $Y=0.64$. Black contour lines indicate increments by $0.01$.}
\label{fig:barcduiduj}
\end{figure}
Further insight on the local structure of turbulence in the detachment zone is obtained by looking at $\aver{\del[v]\del[v]}$ and $\aver{\del[w]\del[w]}$ in the $\left(X, r_x, r_z \right)$ space, shown in figure~\ref{fig:barcduiduj} for $\left(Y, r_y \right) = \left(0.64, 0 \right)$. Identifying spanwise-oriented structures requires considering scales $r_x$ along the inhomogeneous streamwise direction. Indeed $\aver{\del[v]\del[v]}$ locally peaks at $\left(X, r_x, r_z \right) = \left(-0.95,0.3,0 \right)$, i.e. exactly at the $X$ position where the boundary layer detaches and for a specific streamwise scale. This confirms the suggestion by \cite{cimarelli-leonforte-angeli-2018} that spanwise-oriented structures are indeed present. $\aver{\del[w]\del[w]}$ too exhibits a local maximum for finite $r_x$, precisely at $\left( X, r_x, r_z \right) = \left(-1.13, 0.65, 0 \right)$. However, the streamwise extent of this peak is larger than that for $\aver{\del[v]\del[v]}$. Moreover,  $\aver{\del[w]\del[w]}$ increases within the secondary recirculation bubble, where it features a non-monotonic behaviour in $r_z$, while $\aver{\del[v]\del[v]}$ does not. Hence, the detached reverse boundary layer and, in particular, the secondary recirculation bubble appear to be populated by a broader range of structures than just spanwise-oriented vortices. 

\begin{figure}
\centering
\includegraphics[width=0.495\textwidth,trim = 40mm 0mm 40mm 0mm, clip]{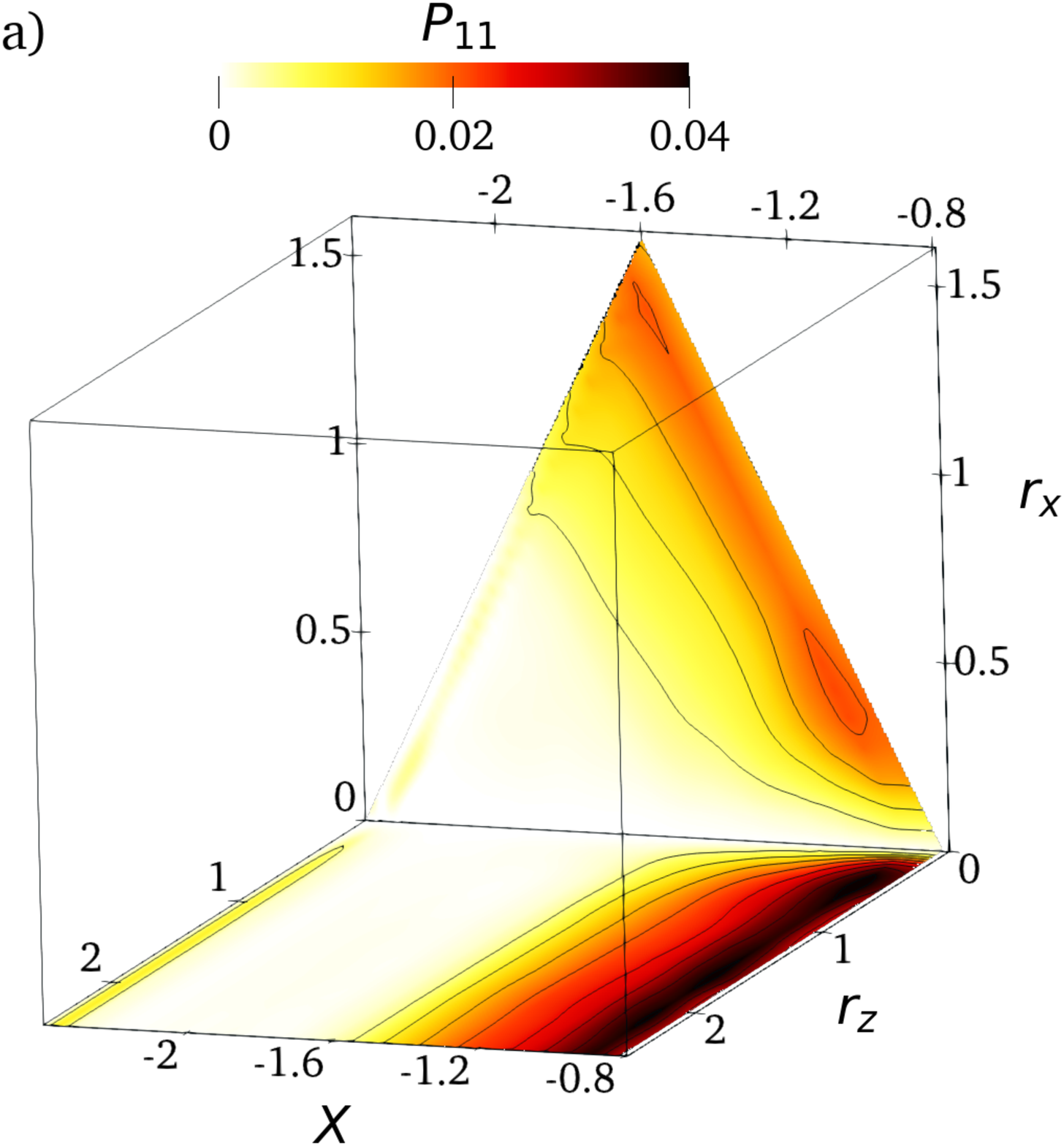}
\includegraphics[width=0.495\textwidth,trim = 40mm 0mm 40mm 0mm, clip]{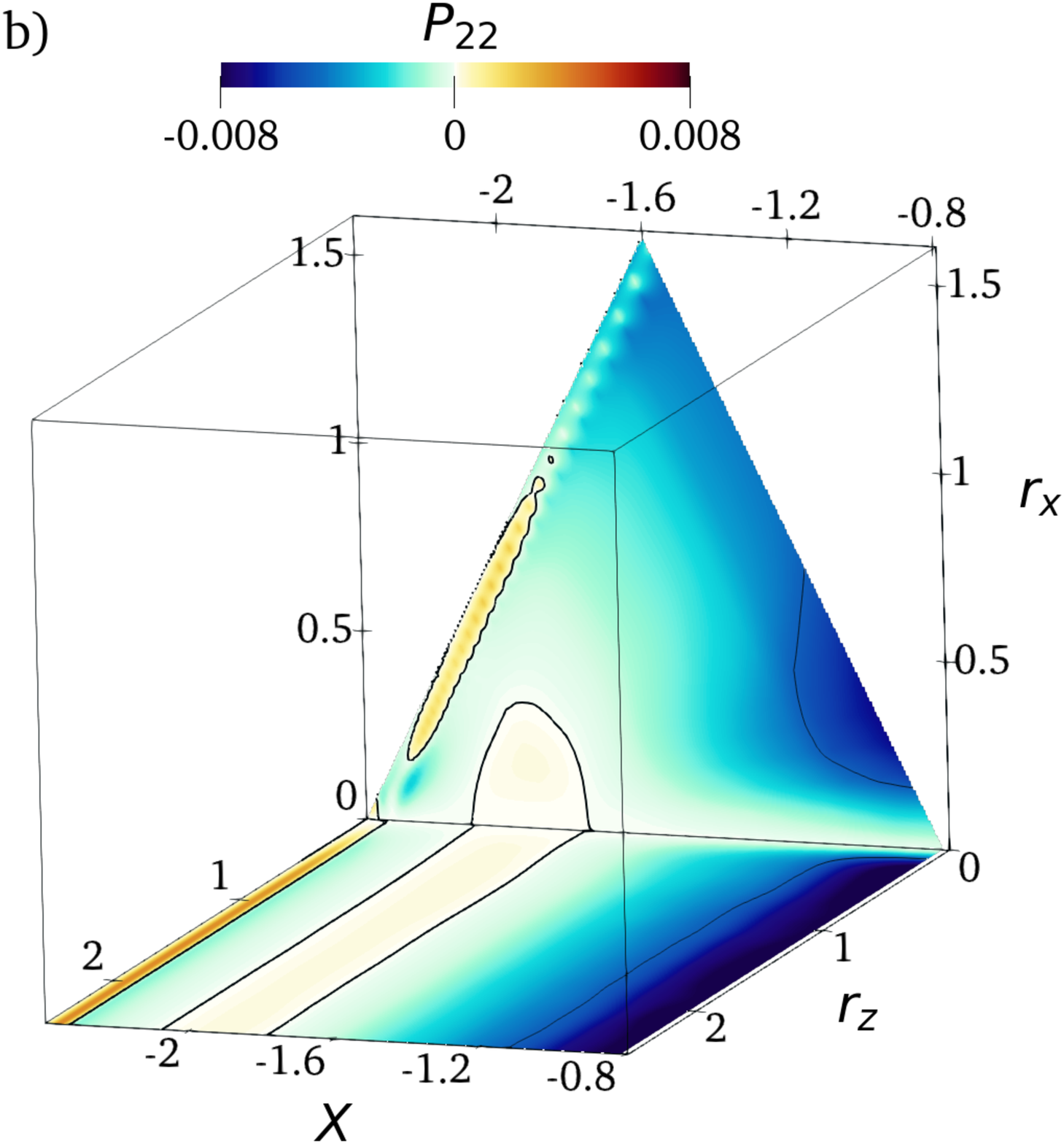}
\caption{Production terms $P_{11}$ (a) and $P_{22}$ (b) in the $(X,r_x,r_z)$-space for $r_y=0$ and $Y=0.64$. Black contour lines indicate increments by $0.005$.}
\label{fig:barcprod}
\end{figure}
The process behind the formation of spanwise-aligned structures is addressed in figure~\ref{fig:barcprod}, which shows the production terms $P_{11}$ and $P_{22}$ in the same $\left(X, r_x, r_z \right)$ space of figure~\ref{fig:barcduiduj}. $P_{11}$ has a local maximum at $\left(X, r_x, r_z \right) = \left(-1.05,0.35,0 \right)$. At these scales, the streamwise fluctuations drain energy from the mean shear and feed $\aver{\del[v] \del[v]}$, which has been connected at such scales to spanwise structures. The process is described by the pressure-strain terms: at these scales indeed it is found (not shown) that $\Pi_{11}<0$ and $\Pi_{22}>0$. 
Similarly, $P_{22}$ is negative everywhere, with a relative minimum in same range of scales where $P_{11}$ is maximum. Thus, $P_{22}$ reconverts the energy $\aver{\del[v] \del[v]}$ received via pressure strain back to the mean flow, thereby sustaining the detachment of the reverse boundary layer. 

Hence, within the limits of this necessarily brief example, the AGKE successfully confirm the literature suggestion that spanwise-oriented structures exist at the detachment of the reverse boundary layer. Moreover, they reveal that these structures do not simply derive from the upstream streamwise-oriented ones simply via a gradual reorientation. Instead, their appearance is rather abrupt, mediated by pressure-strain redistribution but mainly driven by local positive and negative production.

\section{Concluding discussion}
\label{sec:conclusions}

Exact budget equations for the components of the second-order structure function tensor $\aver{ \delta u_i \delta u_j } (\bm{X}, \bm{r}) $ have been considered.
Because of its close relationship with two-point velocity correlations and spectra, $\aver{\duiduj}$ is interpreted as scale Reynolds stress. In this spirit, the budget equations, that we name Anisotropic Generalized Kolmogorov Equations (AGKE), describe production, transport and dissipation of the scale Reynolds stresses in the combined physical and scale space.

Compared to the Generalized Kolmogorov Equation (GKE), which is half the trace of the AGKE and thus describes scale energy only, the AGKE fully account for the anisotropy of the structure function tensor, and allow the description of purely redistributive processes like pressure-strain. They are a powerful tool to complement energy spectra of turbulent fluctuations and spectral Reynolds stress budgets \citep[see, for instance,][]{mizuno-2016,lee-moser-2019}, to which they add two major features: i) scales are defined along directions of statistical inhomogeneity; and ii) fluxes are defined in the space of scales. Thanks to the former feature, scale properties of turbulence can be assessed also along the wall-normal direction of wall-bounded turbulent flows and, in general, in complex turbulent flows. Thanks to the second feature, fluxes of $\aver{\duiduj}$ across all scales and in physical space can be clearly recognised. Thus, beside the identification of scales acting as donors or receivers of scale Reynolds stresses, already possible within the framework of spectral Reynolds stress budgets, the AGKE allow to quantify the local direction of the fluxes of $\aver{\duiduj}$ throughout the whole $\left(\bm{X}, \bm{r}\right)$ space, informing on the different  physical processes underlying the transfer of scale Reynolds stress in space or through scales at different spatial positions in the flow. 

The AGKE have been demonstrated via three examples. With a low-$Re$ turbulent plane channel flow, the near-wall turbulence cycle has been observed and described in terms of the AGKE, thanks to its multi-dimensional and multi-component information. The pressure-strain term of the diagonal components of $\aver{\duiduj}$ is analysed to identify scales and positions involved in the inter-component energy redistribution processes. 
Moreover, the budget equation for the off-diagonal component $\aver{-\del[u]\del[v]}$, the other important element that the AGKE adds to the GKE, is presented and discussed. In contrast to the energetic interpretation of the diagonal components, the scale Reynolds shear stress $\aver{-\del[u]\del[v]}$ is not positive definite, and is rather interpreted as statistical proxy for coherent structures and related to the production of $\aver{\del[u]\del[u]}$. The main transport mechanisms are identified via the combined analysis of the AGKE terms and of the correlation levels along typical transport patterns in the physical and scale space.

Channel flows at higher $Re$ (up to $Re_\tau=1000$) are also considered in order to demonstrate the AGKE on flows characterised by a broader range of scales with particular focus on the outer cycle of wall-turbulence. The range of scales and positions responsible for the net production of streamwise turbulent fluctuations in the outer layer are identified. In particular, the presence of two well-separated self-regenerating cycles belonging to scales attached to the wall and to very-large scale motions are unequivocally detected in a quantitative way.

Finally, the separating and reattaching flow over a finite rectangular cylinder is considered as a test case with two inhomogeneous directions. The AGKE describe how streamwise-oriented structures in the reverse boundary layer within the main recirculation bubble become spanwise-oriented structures in the detachment region. The pressure-strain and production terms show that the spanwise structures form abruptly near the detachment, rather than being gradually reoriented.

The AGKE are a tool with several potential applications. Thanks to the relationship between $\duiduj$ and the unresolved stresses \citep{cimarelli-abba-germano-2019}, the AGKE can be useful to develop large-eddy turbulence models. Indeed, \cite{cimarelli-deangelis-2014} already used the GKE {\em a posteriori} to improve modeling, and the AGKE could further this approach, by fully accounting for anisotropy, an essential property of wall-bounded turbulent flows. For canonical turbulent flows at large values of $Re$, the AGKE seem apt to comprehensively describe the large-scale structures involved in the outer regeneration cycle~\citep{hwang-cossu-2010} and  their modulating effect~\citep{mathis-hutchins-marusic-2009} onto near-wall turbulence.  Such structures, characterised by a large wall-normal extent \citep{hutchins-marusic-2007}, may be involved in a non-negligible transfer of $\aver{\duiduj}$ across wall-normal scales, which is captured by the AGKE but escapes either the spectral Reynolds stress budgets and the analysis based upon structure function alone \citep{agostini-leschziner-2017}. Similarly, in plane Couette flow the AGKE could be used to study the transfer from small to large scales, resulting from the interaction of small near-wall structures with large scales further from the wall, which has been experimentally observed by \cite{kawata-alfredsson-2018} only for the Reynolds shear stress but not for the normal components. The AGKE can also be used to study how turbulent wall-bounded flows are modified by drag reduction \citep{chiarini-quadrio-gatti-2019}. 

Beside their application to canonical flows, the present paper demonstrates that AGKE can provide significant contributions in the study of all those complex flows, such as a backward-facing step, a three-dimensional turbulent boundary layer, flows over complex surfaces, with shear layers and with separation, where anisotropy and inhomogeneity are important.

\section*{Acknowledgments}
D.G. gratefully acknowledges the financial support of the German Research Foundation (DFG) through the priority programme SPP1881 ``Turbulent Superstructures''. Computing time has been provided by the computational resource ForHLR Phase I funded by the Ministry of Science, Research and the Arts, Baden-Württemberg and DFG. 

\section*{Declaration of interests}
The authors report no conflict of interest

\appendix
\section{Budget equation for $\aver{\delta u_i \delta u_j}$}
\label{sec:derivation}

This appendix documents the derivation of the AGKE, following the approach of \cite{danaila-etal-2001}, and reports the complete form for every component of $\aver{ \delta u_i \delta u_j}$.

Let us consider two independent points within the flow domain, $\bm{x}$ and $\bm{x'}$, separated by the increment $\bm{r}=\bm{x'}-\bm{x}$. All quantities in $\bm{x'}$ are denoted with the superscript $'$. $v_i$ ($v'_i$) and $p$ ($p'$) are the velocity components and pressure at $x_i$ ($x'_i$). The incompressible Navier--Stokes equations are written at the two points:
%
\begin{equation}
\frac{\partial v_i}{\partial t} + v_k \frac{\partial v_i}{\partial x_k} = -\frac{1}{\rho} \frac{\partial p}{\partial x_i} + \nu \frac{\partial^2 v_i}{\partial x_k \partial x_k},
\end{equation}
\begin{equation}
\frac{\partial v'_i}{\partial t} + v'_k \frac{\partial v'_i }{\partial x'_k} = -\frac{1}{\rho} \frac{\partial p'}{\partial x'_i} + \nu \frac{\partial^2 v'_i}{\partial x'_k \partial x'_k }.
\end{equation}
where $\rho$ is the fluid density, and $p$ the pressure.

The two points $\bm{x}$ and $\bm{x'}$ are independent: hence $v_i$ and $p$ only depend on $\bm{x}$, while $v'_i$ and $p'$ only depend on $\bm{x'}$, and
\begin{equation}
\frac{\partial v_i}{\partial x'_k}=0,  \ \ \ \ \frac{\partial p}{\partial x'_k}=0 ;
\end{equation}
\begin{equation}
\frac{\partial v'_i}{\partial x_k }=0, \ \ \ \ \frac{\partial p'}{\partial x_k}=0.
\end{equation}
The Reynolds decomposition of the velocity field is now introduced: $v_i=U_i+u_i$ where $U_i=\aver{v_i}$ denotes the mean velocity and $u_i$ the fluctuations. The two equations become:
\begin{equation}
\frac{\partial U_i}{\partial t} + \frac{\partial u_i}{\partial t} + U_k \frac{\partial U_i}{\partial x_k} + U_k \frac{\partial u_i}{\partial x_k} + u_k \frac{\partial  U_i}{\partial x_k} + u_k \frac{\partial u_i}{\partial x_k} = -\frac{1}{\rho} \frac{\partial p}{\partial x_i} + \nu \frac{\partial^2 U_i}{\partial x_k \partial x_k} + \nu \frac{\partial^2 u_i}{\partial x_k \partial x_k}
\label{Re_dec_fix}
\end{equation}
\begin{equation}
\frac{\partial U'_i}{\partial t} + \frac{\partial u'_i}{\partial t} + U'_k \frac{\partial U'_i}{\partial x'_k} + U'_k \frac{\partial u'_i}{\partial x'_k} + u'_k \frac{\partial U'_i}{\partial x'_k} + u'_k \frac{\partial u'_i}{\partial x'_k}  = -\frac{1}{\rho} \frac{\partial p'}{\partial x'_i} + \nu \frac{\partial^2 U'_i}{\partial x'_k \partial x'_k} + \nu \frac{\partial^2 u'_i}{\partial x'_k \partial x'_k}.
\label{Re_dec_mov}
\end{equation}
By subtracting equation \eqref{Re_dec_fix} from \eqref{Re_dec_mov} and using the following relations, derived from the independence of $\bm{x}$ and $\bm{x'}$,
\begin{equation*}
u'_k \frac{\partial U'_i}{\partial x'_k} - u_k \frac{\partial U_i}{\partial x_k} = u'_k \frac{\partial \delta U_i}{\partial x'_k} + u_k \frac{\partial \delta U_i}{\partial x_k}
\end{equation*}
\begin{equation*}
- \frac{1}{\rho} \frac{\partial p'}{\partial x'_i} + \frac{1}{\rho} \frac{\partial p}{\partial x_i} = - \frac{1}{\rho} \left( \frac{\partial}{\partial x'_i} + \frac{\partial}{\partial x_i} \right) \delta p
\end{equation*}
\begin{equation*}
\nu \frac{\partial^2 u'_i}{\partial x'_k \partial x'_k} - \nu \frac{\partial^2 u_i}{\partial x_k \partial x_k} = \nu \left( \frac{\partial^2}{\partial x'_k \partial x'_k} + \frac{\partial^2}{\partial x_k \partial x_k} \right) \delta u_i
\end{equation*}
an equation for the velocity increment $\delta u_i=u'_i-u_i$ is obtained:
\begin{equation}
\begin{gathered}
\frac{\partial \delta U_i}{\partial t} + \frac{\partial \delta u_i}{\partial t} + U'_k \frac{\partial \delta U_i}{\partial x'_k} + U_k \frac{\partial \delta U_i}{\partial x_k} + U'_k \frac{\partial \delta u_i }{\partial x'_k} + U_k \frac{\partial \delta u_i}{\partial x_k}+\\
+ u'_k \frac{\partial \delta U_i}{\partial x'_k} + u_k \frac{\partial \delta U_i}{\partial x_k} + u'_k \frac{\partial \delta u_i}{\partial x'_k} + u_k \frac{\partial \delta u_i}{\partial x_k}=\\
- \frac{1}{\rho} \left( \frac{\partial}{\partial x'_i} + \frac{\partial}{\partial x_i} \right) \delta p + \nu \left( \frac{\partial^2}{\partial x'_k \partial x'_k} + \frac{\partial^2}{ \partial x_k \partial x_k} \right) \delta U_i + \nu \left( \frac{\partial^2}{\partial x'_k \partial x'_k} + \frac{\partial^2}{\partial x_k \partial x_k} \right) \delta u_i.
\end{gathered}
\label{eq:3}
\end{equation}
By adding and subtracting $u_k \left( {\partial \delta u_i}/{\partial x'_k} \right) + u_k \left( {\partial \delta U_i}/{\partial x'_k} \right) + U_k \left( {\partial \delta u_i}/{\partial x'_k} \right) + U_k \left( {\partial \delta U_i}/{\partial x'_k} \right)$ to the left-hand side and observing that
%
\begin{equation*}
u'_k \frac{\partial \delta U_i}{\partial x'_k} = \delta u_k \frac{\partial \delta U_i}{\partial x'_k} + u_k \frac{\partial \delta U_i}{\partial x'_k} ,
\end{equation*}
%
%
equation \eqref{eq:3} becomes
\begin{equation}
\begin{gathered}
\frac{\partial \delta U_i}{\partial t} + \frac{\partial \delta u_i}{\partial t} + \delta U_k \frac{\partial \delta U_i}{\partial x'_k} + U_k \left( \frac{\partial}{\partial x'_k} + \frac{\partial}{\partial x_k} \right) \delta U_i+ \delta U_k \frac{\partial \delta u_i}{\partial x'_k} + U_k \left( \frac{\partial}{\partial x'_k} + \frac{\partial}{\partial x_k} \right) \delta u_i + \\
+\delta u_k \frac{\partial \delta U_i}{\partial x'_k} + u_k \left( \frac{\partial}{\partial x'_k} + \frac{\partial}{\partial x_k} \right) \delta U_i+ \delta u_k \frac{\partial \delta u_i}{\partial x'_k} + u_k \left( \frac{\partial}{\partial x'_k} + \frac{\partial}{\partial x_k} \right) \delta u_i=\\
- \frac{1}{\rho} \left( \frac{\partial}{\partial x'_i} + \frac{\partial}{\partial x_i} \right) \delta p + \nu \left( \frac{\partial^2}{\partial x'_k \partial x'_k} + \frac{\partial^2}{\partial x_k \partial x_k} \right) \delta U_i + \nu \left( \frac{\partial^2}{\partial x'_k \partial x'_k} + \frac{\partial^2}{\partial x_k \partial x_k} \right) \delta u_i.
\end{gathered}
\label{eq:4}
\end{equation}
%
%
Equation \eqref{eq:4} multiplied by $\delta u_j$ is now summed to the same equation, with the $i$-index switched to $j$-index and after multiplication by $\delta u_i$. We then use incompressibility and again independence of $\bm{x}$ and $\bm{x'}$ to obtain:
\begin{equation}
\begin{gathered}
\delta u_j \frac{\partial \delta U_i}{\partial t} + \delta u_i \frac{\partial \delta U_j}{\partial t} + \frac{\partial}{\partial t} \left( \delta u_i \delta u_j \right)
+ \delta u_j \delta U_k \frac{\partial \delta U_i}{\partial x'_k} + \delta u_i \delta U_k \frac{\partial \delta U_j}{\partial x'_k}
+ \delta u_j U_k \left( \frac{\partial}{\partial x'_k} + \frac{\partial}{\partial x_k} \right) \delta U_i +\\
+ \delta u_i U_k \left( \frac{\partial}{\partial x'_k} + \frac{\partial}{\partial x_k} \right) \delta U_j
+ \frac{\partial}{\partial x'_k} \left( \delta U_k \delta u_i \delta u_j \right) 
+ \left( \frac{\partial}{\partial x'_k} + \frac{\partial}{\partial x_k} \right) \left( U_k \delta u_i \delta u_j \right)+\\
+ \delta u_j \delta u_k \frac{\partial \delta U_i}{\partial x'_k} + \delta u_i \delta u_k \frac{\partial \delta U_j}{\partial x'_k}
+ \delta u_j u_k \left( \frac{\partial}{\partial x'_k} + \frac{\partial}{\partial x_k} \right) \delta U_i + \delta u_i u_k \left( \frac{\partial}{\partial x'_k} + \frac{\partial}{\partial x_k} \right) \delta U_j+\\
+ \frac{\partial}{\partial x'_k} \left( \delta u_k \delta u_i \delta u_j \right) 
+ \left( \frac{\partial}{\partial x'_k} + \frac{\partial}{\partial x_k} \right) \left( u_k \delta u_i \delta u_j \right)=\\
- \frac{1}{\rho} \left[ \left( \frac{\partial}{\partial x'_i} + \frac{\partial}{\partial x_i} \right) \left( \delta p \delta u_j \right) + \left( \frac{\partial}{\partial x'_j} + \frac{\partial}{\partial x_j} \right) \left( \delta p \delta u_i \right) \right]+\\
+ \frac{1}{\rho} \left[ \delta p \left( \frac{\partial}{\partial x'_i} + \frac{\partial}{\partial x_i} \right) \delta u_j + \delta p \left( \frac{\partial}{\partial x'_j} + \frac{\partial}{\partial x_j} \right) \delta u_i \right]+\\
+ \nu \left[ \delta u_j \left( \frac{\partial^2}{\partial x'_k \partial x'_k} + \frac{\partial^2}{\partial x_k \partial x_k} \right) \left( \delta U_i + \delta u_i \right) \right] + \nu \left[ \delta u_i \left( \frac{\partial^2}{\partial x'_k \partial x'_k} + \frac{\partial}{\partial x_k \partial x_k} \right) \left( \delta U_j + \delta u_j \right) \right]
\end{gathered}
\label{eq:6}
\end{equation}
The averaging operator is now applied:
\begin{equation}
\begin{gathered}
\frac{\partial}{\partial t} \aver{\delta u_i \delta u_j} + \frac{\partial}{\partial x'_k} \aver{\delta U_k \delta u_i \delta u_j}
+ \left( \frac{\partial}{\partial x'_k} + \frac{\partial}{\partial x_k} \right) \aver{U_k \delta u_i \delta u_j}+\\
+ \aver{\delta u_j \delta u_k} \frac{\partial \delta U_i}{\partial x'_k} + \aver{\delta u_i \delta u_k} \frac{\partial \delta U_j}{\partial x'_k}+\\
+ \aver{\delta u_j u_k} \left( \frac{\partial}{\partial x'_k} + \frac{\partial}{\partial x_k} \right) \delta U_i + \aver{\delta u_i u_k} \left( \frac{\partial}{\partial x'_k} + \frac{\partial}{\partial x_k} \right) \delta U_j+\\
+ \frac{\partial}{\partial x'_k}\aver{\delta u_k \delta u_i \delta u_j} + \left( \frac{\partial}{\partial x'_k} + \frac{\partial}{\partial x_k} \right)\aver{u_k \delta u_i \delta u_j}=\\
- \frac{1}{\rho} \left[ \left( \frac{\partial}{\partial x'_i} + \frac{\partial}{\partial x_i} \right) \aver{\delta p \delta u_j} + \left( \frac{\partial}{\partial x'_j}+ \frac{\partial}{\partial x_j} \right) \aver{\delta p \delta u_i} \right]+\\
+ \frac{1}{\rho} \aver{\delta p \left( \frac{\partial}{\partial x'_i} + \frac{\partial}{\partial x_i} \right) \delta u_j} + \frac{1}{\rho} \aver{\delta p \left( \frac{\partial}{\partial x'_j} + \frac{\partial}{\partial x_j} \right) \delta u_i}+\\
+ \nu \aver{\delta u_j \left( \frac{\partial^2}{\partial x'_k \partial x'_k} + \frac{\partial^2}{\partial x_k \partial x_k} \right) \delta u_i} + \nu \aver{\delta u_i \left( \frac{\partial^2}{\partial x'_k \partial x'_k} + \frac{\partial^2}{\partial x_k \partial x_k} \right) \delta u_j}.
\end{gathered}
\label{eq:7}
\end{equation}
We now introduce a new set of independent variables, $\bm{\Xc}$ and $\bm{\R}$
\begin{equation*}
\Xc_i = \frac{x'_i + x_i}{2}
\end{equation*}
\begin{equation*}
\R_i=x'_i-x_i.
\end{equation*}
The derivatives with respect to $\Xc_i$ and $\R_i$ are related to those with respect to $x'_i$ and $x_i$ as follows:
\begin{equation*}
\frac{\partial}{\partial x_i} = \frac{\partial}{\partial \Xc_i} \frac{\partial \Xc_i}{\partial x_i} + \frac{\partial}{\partial \R_i} \frac{\partial \R_i}{\partial x_i} = \frac{1}{2} \frac{\partial}{\partial \Xc_i} - \frac{\partial}{\partial \R_i}
\end{equation*}
\begin{equation*}
\frac{\partial}{\partial x'_i} = \frac{\partial}{\partial \Xc_i} \frac{\partial \Xc_i}{\partial x'_i} + \frac{\partial}{\partial \R_i} \frac{\partial \R_i}{\partial x'_i} = \frac{1}{2} \frac{\partial}{\partial \Xc_i} + \frac{\partial}{\partial \R_i}
\end{equation*}
\begin{equation*}
\frac{\partial^2}{\partial x'_k \partial x'_k} + \frac{\partial^2}{\partial x_k \partial x_k} = \frac{1}{2} \frac{\partial^2}{\partial \Xc_k \partial \Xc_k} + 2 \frac{\partial^2}{\partial \R_k \partial \R_k}.
\end{equation*}
By using $\bm{\Xc}$ and $\bm{\R}$ as independent variables, equation \eqref{eq:7} can be further rewritten as:
\begin{equation}
\begin{gathered}
\frac{\partial}{\partial t} \aver{\duiduj} + \left( \frac{1}{2} \frac{\partial}{\partial \Xc_k} + \frac{\partial}{\partial \R_k} \right) \aver{\delta U_k \delta u_i \delta u_j}
+ \frac{\partial}{\partial \Xc_k} \aver{U_k \delta u_i \delta u_j}
+ \aver{\delta u_j \delta u_k} \left( \frac{1}{2} \frac{\partial}{\partial \Xc_k} + \frac{\partial}{\partial \R_k} \right) \delta U_i + \\
+ \aver{\delta u_i \delta u_k} \left( \frac{1}{2} \frac{\partial}{\partial \Xc_k} + \frac{\partial}{\partial \R_k} \right) \delta U_j
+ \aver{\delta u_j u_k} \frac{\partial \delta U_i}{\partial \Xc_k} + \aver{\delta u_i u_k} \frac{\partial \delta U_j}{\partial \Xc_k}+\\
+ \left( \frac{1}{2} \frac{\partial}{\partial \Xc_k} + \frac{\partial}{\partial \R_k} \right) \aver{\delta u_k \delta u_i \delta u_j}
+ \frac{\partial}{\partial \Xc_k}\aver{u_k \delta u_i \delta u_j}=\\
- \frac{1}{\rho} \left( \frac{\partial}{\partial \Xc_i} \aver{\delta p \delta u_j} + \frac{\partial}{\partial \Xc_j} \aver{\delta p \delta u_i} \right)
+ \frac{1}{\rho} \aver{\delta p \frac{\partial \delta u_j}{\partial \Xc_i}} + \frac{1}{\rho} \aver{\delta p \frac{\partial \delta u_i}{\partial \Xc_j}}+\\
+\nu \aver{ \delta u_j \left( \frac{\partial^2}{\partial x'_k \partial x'_k} + \frac{\partial^2}{\partial x_k \partial x_k} \right) \delta u_i}
+\nu \aver{ \delta u_i \left( \frac{\partial^2}{\partial x'_k \partial x'_k} + \frac{\partial^2}{\partial x_k \partial x_k} \right) \delta u_j}.
\end{gathered}
\label{eq:8}
\end{equation}
The viscous term can be written more compactly as:
\begin{equation*}
\begin{gathered}	
\nu \aver{\delta u_j \left( \frac{\partial^2}{\partial x'_k \partial x'_k} + \frac{\partial^2}{\partial x_k \partial x_k} \right) \delta u_i}
+\nu \aver{\delta u_i \left( \frac{\partial^2}{\partial x'_k \partial x'_k} + \frac{\partial^2}{\partial x_k \partial x_k} \right) \delta u_j}= \\
\frac{\nu}{2} \frac{\partial^2}{\partial \Xc_k \partial \Xc_k} \aver{\delta u_i \delta u_j} +
2 \nu \frac{\partial^2}{\partial \R_k \partial \R_k} \aver{\delta u_i \delta u_j} 
- \nu \aver{\frac{\partial \delta u_i}{\partial \Xc_k} \frac{\partial \delta u_j}{\partial \Xc_k}} -
4 \nu \aver{\frac{\partial \delta u_i}{\partial \R_k} \frac{\partial \delta u_j}{\partial \R_k}} =\\
\frac{\nu}{2} \frac{\partial^2}{\partial \Xc_k \partial \Xc_k} \aver{\delta u_i \delta u_j}
+2 \nu \frac{\partial^2}{\partial \R_k \partial \R_k} \aver{\delta u_i \delta u_j}
-2\left( \epsilon'_{ij} + \epsilon_{ij} \right)
\end{gathered}
\end{equation*}
where:
\begin{equation*}
\epsilon_{ij}=\nu \aver{\frac{\partial u_i}{\partial x_k} \frac{\partial u_j}{\partial x_k} }.
\end{equation*}
Finally, by using  in Eq. \eqref{eq:8} the following relations
\begin{equation*}
\frac{1}{2} \frac{\partial}{\partial \Xc_k} \aver{\delta U_k \delta u_i \delta u_j} + \frac{\partial}{\partial \Xc_k}\aver{U_k \delta u_i \delta u_j}=\frac{\partial}{\partial \Xc_k}\aver{U_k^* \delta u_i \delta u_j}
\end{equation*}
\begin{equation*}
\aver{\delta u_j \delta u_k} \left( \frac{1}{2} \frac{\partial}{\partial \Xc_k} + \frac{\partial}{\partial \R_k} \right) \delta U_i + \aver{\delta u_j u_k} \frac{\partial \delta U_i}{\partial \Xc_k}=
\aver{\delta u_j u_k^*} \delta \left( \frac{\partial U_i}{\partial x_k} \right) + \aver{\delta u_j \delta u_k} \left( \frac{\partial U_i}{\partial x_k} \right)^*
\end{equation*}
where the superscript $*$ denotes the average of a generic quantity $f$ at positions $\bm{\Xc} \pm \bm{\R}/2$:

\begin{equation*}
\mid{f}=\frac{ f(\bm{\Xc}+\bm{\R}/2) + f(\bm{\Xc}-\bm{\R}/2)}{2}
\end{equation*}
one arrives at the final form of the AGKE:
\begin{equation}
\begin{gathered}
\frac{\partial}{\partial t} \aver{\duiduj} + \frac{\partial}{\partial \R_k}\aver{\delta U_k \delta u_i \delta u_j}+\frac{\partial}{\partial \R_k}\aver{\delta u_k \delta u_i \delta u_j}-2 \nu \frac{\partial^2}{\partial \R_k \partial \R_k} \aver{\delta u_i \delta u_j}+\frac{\partial}{\partial \Xc_k}\aver{U_k^* \delta u_i \delta u_j} +\\
+ \frac{\partial}{\partial \Xc_k}\aver{u_k^* \delta u_i \delta u_j}+\frac{1}{\rho}\left( \frac{\partial}{\partial \Xc_j}\aver{\delta p \delta u_i} + \frac{\partial}{\partial \Xc_i}\aver{\delta p \delta u_j} \right)-\frac{\nu}{2}\frac{\partial^2}{\partial \Xc_k \partial \Xc_k}\aver{\delta u_i \delta u_j}=\\
-\aver{u_k^* \delta u_j} \delta \left( \frac{\partial U_i}{\partial x_k} \right)-\aver{u_k^* \delta u_i} \delta \left( \frac{\partial U_j}{\partial x_k} \right)-\aver{\delta u_k \delta u_j} \left( \frac{\partial U_i}{\partial x_k} \right)^*-\aver{\delta u_k \delta u_i} \left( \frac{\partial U_j}{\partial x_k} \right)^*+\\
+\frac{1}{\rho}\aver{\delta p \frac{\partial \delta u_i}{\partial \Xc_j}}+\frac{1}{\rho}\aver{\delta p \frac{\partial \delta u_j}{\partial \Xc_i}}-4 \epsilon_{ij}^*.
\end{gathered}
\label{eq:9}
\end{equation}
The AGKE can be written in divergence form
\begin{equation}
\frac{\partial \aver{\duiduj}}{\partial t} + \deriv{\phir{k}{ij}}{\R_k} + \deriv{\phic{k}{ij}}{\Xc_k} = \src[ij]
\end{equation}
where $\phir{k}{ij}$ and $\phic{k}{ij}$ are the components in the space of scales $\R_k$ and in the physical space $\Xc_k$ of the six dimensional vector field of fluxes $\bm{\Phi}_{ij}=(\bm{\phi}_{ij},\bm{\psi}_{ij})$, and $\src[ij]$ is the source term. These tensor are defined by the expressions below, where $\delta_{ij}$ denotes the Kroenecker delta:
\begin{equation*}
\phir{k}{ij}=\aver{\delta U_k \delta u_i \delta u_j}+\aver{\delta u_k \delta u_i \delta u_j} -2 \nu \frac{\partial}{\partial \R_k} \aver{\delta u_i \delta u_j} \ \ \ k=1,2,3
\end{equation*}
\begin{equation*}
\phic{k}{ij}=\aver{U_k^* \delta u_i \delta u_j} + \aver{u_k^* \delta u_i \delta u_j} +\frac{1}{\rho} \aver{\delta p \delta u_i} \delta_{kj} + \frac{1}{\rho} \aver{\delta p \delta u_j} \delta_{ki} - \frac{\nu}{2} \frac{\partial}{\partial \Xc_k} \aver{\delta u_i \delta u_j} \ \ \ k=1,2,3
\end{equation*}
\begin{equation*}
\begin{split}
\src[ij] =
&-\aver{v_k^* \delta u_j} \delta \left( \frac{\partial U_i}{\partial x_k} \right)-\aver{v_k^* \delta u_i} \delta \left( \frac{\partial U_j}{\partial x_k} \right)
-\aver{\delta u_k \delta u_j} \left( \frac{\partial U_i}{\partial x_k} \right)^*-\aver{\delta u_k \delta u_i} \left( \frac{\partial U_j}{\partial x_k} \right)^*+\\
&+\frac{1}{\rho}\aver{\delta p \frac{\partial \delta u_i}{\partial \Xc_j}}+\frac{1}{\rho}\aver{\delta p \frac{\partial \delta u_j}{\partial \Xc_i}}-4 \mid{\epsilon_{ij}}.
\end{split}
\end{equation*}

The six complete AGKE components are reported below.

\subsection{$\aver{\delta u_1 \delta u_1}$}

\begin{equation}
\begin{gathered}
\frac{\partial}{\partial t} \aver{\delta u_1 \delta u_1} + \frac{\partial}{\partial \R_k}\aver{\delta U_k \delta u_1 \delta u_1}+\frac{\partial}{\partial \R_k}\aver{\delta u_k \delta u_1 \delta u_1}-2 \nu \frac{\partial^2}{\partial \R_k \partial \R_k} \aver{\delta u_1 \delta u_1}+\\
+\frac{\partial}{\partial \Xc_k}\aver{U_k^* \delta u_1 \delta u_1} + \frac{\partial}{\partial \Xc_k}\aver{u_k^* \delta u_1 \delta u_1}+\frac{2}{\rho} \frac{\partial}{\partial \Xc_1}\aver{\delta p \delta u_1} -\frac{\nu}{2}\frac{\partial^2}{\partial \Xc_k \partial \Xc_k}\aver{\delta u_1 \delta u_1}=\\
-2\aver{u_k^* \delta u_1} \delta \left( \frac{\partial U_1}{\partial x_k} \right)-2\aver{\delta u_k \delta u_1} \left( \frac{\partial U_1}{\partial x_k} \right)^*+\frac{2}{\rho}\aver{\delta p \frac{\partial \delta u_1}{\partial \Xc_1}}-4 \mid{\epsilon_{11}}
\end{gathered}
\label{eq:AGKE_uu}
\end{equation}

\subsection{$\aver{\delta u_2 \delta u_2}$}

\begin{equation}
\begin{gathered}
\frac{\partial}{\partial t} \aver{\delta u_2 \delta u_2} + \frac{\partial}{\partial \R_k}\aver{\delta U_k \delta u_2 \delta u_2}+\frac{\partial}{\partial \R_k}\aver{\delta u_k \delta u_2 \delta u_2}-2 \nu \frac{\partial^2}{\partial \R_k \partial \R_k} \aver{\delta u_2 \delta u_2}+\\
+\frac{\partial}{\partial \Xc_k}\aver{U_k^* \delta u_2 \delta u_2} + \frac{\partial}{\partial \Xc_k}\aver{u_k^* \delta u_2 \delta u_2}+\frac{2}{\rho} \frac{\partial}{\partial \Xc_2}\aver{\delta p \delta u_2} -\frac{\nu}{2}\frac{\partial^2}{\partial \Xc_k \partial \Xc_k}\aver{\delta u_2 \delta u_2}=\\
-2\aver{u_k^* \delta u_2} \delta \left( \frac{\partial U_2}{\partial x_k} \right)-2\aver{\delta u_k \delta u_2} \left( \frac{\partial U_2}{\partial x_k} \right)^*+\frac{2}{\rho}\aver{\delta p \frac{\partial \delta u_2}{\partial \Xc_2}}- 4 \mid{\epsilon_{22}}
\end{gathered}
\label{eq:AGKE_vv}
\end{equation}

\subsection{$\aver{\delta u_3 \delta u_3}$}

\begin{equation}
\begin{gathered}
\frac{\partial}{\partial t} \aver{\delta u_3 \delta u_3} + \frac{\partial}{\partial \R_k}\aver{\delta U_k \delta u_3 \delta u_3}+\frac{\partial}{\partial \R_k}\aver{\delta u_k \delta u_3 \delta u_3}-2 \nu \frac{\partial^2}{\partial \R_k \partial \R_k} \aver{\delta u_3 \delta u_3}+\\
+\frac{\partial}{\partial \Xc_k}\aver{U_k^* \delta u_3 \delta u_3} + \frac{\partial}{\partial \Xc_k}\aver{u_k^* \delta u_3 \delta u_3}+\frac{2}{\rho} \frac{\partial}{\partial \Xc_3}\aver{\delta p \delta u_3} -\frac{\nu}{2}\frac{\partial^2}{\partial \Xc_k \partial \Xc_k}\aver{\delta u_3 \delta u_3}=\\
-2\aver{u_k^* \delta u_3} \delta \left( \frac{\partial U_3}{\partial x_k} \right)-2\aver{\delta u_k \delta u_3} \left( \frac{\partial U_3}{\partial x_k} \right)^*+\frac{2}{\rho}\aver{\delta p \frac{\partial \delta u_3}{\partial \Xc_3}}- 4 \mid{\epsilon_{33}}
\end{gathered}
\label{eq:AGKE_ww}
\end{equation}

\subsection{$\aver{\delta u_1 \delta u_2}$}

\begin{equation}
\begin{gathered}
\frac{\partial}{\partial t} \aver{\delta u_1 \delta u_2} +\frac{\partial}{\partial \R_k}\aver{\delta U_k \delta u_1 \delta u_2}+\frac{\partial}{\partial \R_k}\aver{\delta u_k \delta u_1 \delta u_2}-2 \nu \frac{\partial^2}{\partial \R_k \partial \R_k} \aver{\delta u_1 \delta u_2}+\frac{\partial}{\partial \Xc_k}\aver{U_k^* \delta u_1 \delta u_2} +\\
+ \frac{\partial}{\partial \Xc_k}\aver{u_k^* \delta u_1 \delta u_2}+\frac{1}{\rho}\left( \frac{\partial}{\partial \Xc_2}\aver{\delta p \delta u_1} + \frac{\partial}{\partial \Xc_1}\aver{\delta p \delta u_2} \right)-\frac{\nu}{2}\frac{\partial^2}{\partial \Xc_k \partial \Xc_k}\aver{\delta u_1 \delta u_2}=\\
-\aver{u_k^* \delta u_2} \delta \left( \frac{\partial U_1}{\partial x_k} \right)-\aver{u_k^* \delta u_1} \delta \left( \frac{\partial U_2}{\partial x_k} \right)-\aver{\delta u_k \delta u_2} \left( \frac{\partial U_1}{\partial x_k} \right)^*-\aver{\delta u_k \delta u_1} \left( \frac{\partial U_2}{\partial x_k} \right)^*+\\
+\frac{1}{\rho}\aver{\delta p \frac{\partial \delta u_1}{\partial \Xc_2}}+\frac{1}{\rho}\aver{\delta p \frac{\partial \delta u_2}{\partial \Xc_1}}- 4 \mid{\epsilon_{12}}
\end{gathered}
\label{eq:AGKE_uv}
\end{equation}

\subsection{$\aver{\delta u_1 \delta u_3}$}

\begin{equation}
\begin{gathered}
\frac{\partial}{\partial t} \aver{\delta u_1 \delta u_3} + \frac{\partial}{\partial \R_k}\aver{\delta U_k \delta u_1 \delta u_3}+\frac{\partial}{\partial \R_k}\aver{\delta u_k \delta u_1 \delta u_3}-2 \nu \frac{\partial^2}{\partial \R_k \partial \R_k} \aver{\delta u_1 \delta u_3}+\frac{\partial}{\partial \Xc_k}\aver{U_k^* \delta u_1 \delta u_3} +\\
+ \frac{\partial}{\partial \Xc_k}\aver{u_k^* \delta u_1 \delta u_3}+\frac{1}{\rho}\left( \frac{\partial}{\partial \Xc_3}\aver{\delta p \delta u_1} + \frac{\partial}{\partial \Xc_1}\aver{\delta p \delta u_3} \right)-\frac{\nu}{2}\frac{\partial^2}{\partial \Xc_k \partial \Xc_k}\aver{\delta u_1 \delta u_3}=\\
-\aver{u_k^* \delta u_3} \delta \left( \frac{\partial U_1}{\partial x_k} \right)-\aver{u_k^* \delta u_1} \delta \left( \frac{\partial U_3}{\partial x_k} \right)-\aver{\delta u_k \delta u_3} \left( \frac{\partial U_1}{\partial x_k} \right)^*-\aver{\delta u_k \delta u_1} \left( \frac{\partial U_3}{\partial x_k} \right)^*+\\
+\frac{1}{\rho}\aver{\delta p \frac{\partial \delta u_1}{\partial \Xc_3}}+\frac{1}{\rho}\aver{\delta p \frac{\partial \delta u_3}{\partial \Xc_1}}-4 \mid{\epsilon_{13}}
\end{gathered}
\label{eq:AGKE_uw}
\end{equation}

\subsection{$\aver{\delta u_2 \delta u_3}$}

\begin{equation}
\begin{gathered}
\frac{\partial}{\partial t} \aver{\delta u_2 \delta u_3} + \frac{\partial}{\partial \R_k}\aver{\delta U_k \delta u_2 \delta u_3}+\frac{\partial}{\partial \R_k}\aver{\delta u_k \delta u_2 \delta u_3}-2 \nu \frac{\partial^2}{\partial \R_k \partial \R_k} \aver{\delta u_2 \delta u_3}+\frac{\partial}{\partial \Xc_k}\aver{U_k^* \delta u_2 \delta u_3} +\\
+ \frac{\partial}{\partial \Xc_k}\aver{u_k^* \delta u_2 \delta u_3}+\frac{1}{\rho}\left( \frac{\partial}{\partial \Xc_3}\aver{\delta p \delta u_2} + \frac{\partial}{\partial \Xc_2}\aver{\delta p \delta u_3} \right)-\frac{\nu}{2}\frac{\partial^2}{\partial \Xc_k \partial \Xc_k}\aver{\delta u_2 \delta u_3}=\\
-\aver{u_k^* \delta u_3} \delta \left( \frac{\partial U_2}{\partial x_k} \right)-\aver{u_k^* \delta u_2} \delta \left( \frac{\partial U_3}{\partial x_k} \right)-\aver{\delta u_k \delta u_3} \left( \frac{\partial U_2}{\partial x_k} \right)^*-\aver{\delta u_k \delta u_2} \left( \frac{\partial U_3}{\partial x_k} \right)^*+\\
+\frac{1}{\rho}\aver{\delta p \frac{\partial \delta u_2}{\partial \Xc_3}}+\frac{1}{\rho}\aver{\delta p \frac{\partial \delta u_3}{\partial \Xc_2}}- 4 \mid{\epsilon_{23}}
\end{gathered}
\label{eq:AGKE_vw}
\end{equation}

\section{Symmetries}
\label{sec:symmetries}

Here the symmetries of the terms of the AGKE in their specialised form tailored to the indefinite plane channel flow are reported. For simplicity sake, the origin of the wall-normal coordinate is shifted to the centreline of the channel. $x$, $y$ and $z$ indicate the streamwise, wall-normal and spanwise directions, with $u$, $v$ and $w$ the corresponding velocity components.

The terms appearing in the budget equations for $\aver{\delta u \delta u}$, $\aver{\delta v \delta v}$ and $\aver{\delta w \delta w}$ possess the same symmetries as those in the GKE for $\aver{\delta u^2}$ \citep[see][]{cimarelli-deangelis-casciola-2013}. In detail, the transformation $\bm{r} \rightarrow -\bm{r}$ leads to $\bm{\phiR} \rightarrow -\bm{\phiR}$, $\phiC \rightarrow \phiC$, $\src[] \rightarrow \src[]$ and $\aver{\delta u_i \delta u_i} \rightarrow \aver{\delta u_i \delta u_i}$. The inversion of the wall-normal coordinate $y$ leads to $Y \rightarrow -Y$, $ \R_y \rightarrow -\R_y$ and $\phiR_x \rightarrow \phiR_x$, $\phiR_y \rightarrow -\phiR_y$, $\phiR_z \rightarrow \phiR_z$, $\phiC \rightarrow -\phiC$, $\xi \rightarrow \xi$ and $\aver{\delta u_i \delta u_i} \rightarrow \aver{\delta u_i \delta u_i}$. The inversion of the spanwise coordinate $z$ leads to $ \R_z \rightarrow -\R_z $ and $\phiR_x \rightarrow \phiR_x$, $\phiR_y \rightarrow \phiR_y$, $\phiR_z \rightarrow -\phiR_z$, $\phiC \rightarrow \phiC$, $\xi \rightarrow \xi$ and $\aver{\delta u_i \delta u_i} \rightarrow \aver{\delta u_i \delta u_i}$.

The terms appearing in the budget equations for the off-diagonal are slightly different: the inversion of $\bm{\R}$ leads to the same symmetries, whereas the inversion of $y$ and $z$ leads to different changes.
In detail, when $y \rightarrow -y$ the terms related to $\aver{\delta u \delta v}$ and $\aver{\delta v \delta w}$ undergo $\phiR_x \rightarrow -\phiR_x$, $\phiR_y \rightarrow \phiR_y$, $ \phiR_z \rightarrow -\phiR_z $, $ \phiC \rightarrow \phiC$, $\xi \rightarrow -\xi$ and $ \aver{\delta u_i \delta u_j} \rightarrow \ - \aver{\delta u_i \delta u_j}$, whereas when $z \rightarrow -z$ the terms of $\aver{\delta u \delta w}$ and $\aver{\delta v \delta w}$ undergo $\phiR_x \rightarrow -\phiR_x$, $\phiR_y \rightarrow -\phiR_y$, $\phiR_z \rightarrow \phiR_z$, $\phiC \rightarrow -\phiC$, $\xi \rightarrow -\xi$ and $\aver{\delta u_i \delta u_j} \rightarrow - \aver{\delta u_i \delta u_j}$.

The above-described symmetries require that some terms of the AGKE are zero in particular regions of the 4-dimensional domain. These requirements are listed below, for each of the components of $\aver{\delta u_i \delta u_j}$.

\begin{itemize}
	\item $\aver{\delta u \delta u}$, $\aver{\delta v \delta v}$, $\aver{\delta w \delta w}$

	\begin{center}
		\begin{align*}
		\phiR_x(Y,0,0,r_z) &= 0  \hspace{2cm}   &\phiR_x(0,0,r_y,r_z) =0&\\    
		\phiR_y(Y,0,0,r_z) &=0                  &\phiR_y(0,r_x,0,r_z) =0&\\
		\phiR_z(Y,r_x,r_y,0) &= 0 \\ 
		\phiC(Y,0,0,0) &= 0   	          &\phiC(0,r_x,0,r_z) =0&\\
                \phiC(0,0,r_y,r_z) &=0 &
		\end{align*}
	\end{center}

	\item $\aver{\delta u \delta v}$

	\begin{center}
		\begin{align*}
		\phiR_x(Y,0,0,r_z) &= 0  \hspace{2cm}   &\phiR_x(0,r_x,0,r_z) =0&\\
		\phiR_y(Y,0,0,r_z) &= 0                 &\phiR_y(0,0,r_y,r_z)= 0&\\ 
		\phiR_z(Y,r_x,r_y,0) &= 0               &\phiR_z(0,r_x,0,r_z) = 0& \\
		\phiR_z(0,0,r_y,r_z) &= 0 \\
                \phiC(Y,0,0,0) &=0 \\
		\xi(0,r_x,0,r_z) &=0                         &\xi(0,0,r_y,r_z)= 0&\\
		\aver{\delta u \delta v}(0,r_x,0,r_z) &= 0   &\aver{\delta u \delta v}(0,0,r_y,r_z) = 0&
		\end{align*}
	\end{center}

	\item  $\aver{\delta u \delta w}$

	\begin{center}
		\begin{align*}
		\phiR_x(Y,r_x,r_y,0) &= 0 \\
		\phiR_y(Y,r_x,r_y,0) &= 0 \hspace{2cm}    &\phiR_y(0,r_x,0,r_z) = 0&\\
		\phiR_y(0,0,r_y,r_z) &= 0  \\
                \phiR_z(Y,0,0,r_z) &= 0                   &\phiR_z(0,0,r_y,r_z) = 0&\\
                \phiC(Y,r_x,r_y,0) &= 0                   &\phiC(Y,0,0,r_z) = 0&\\
                \phiC(0,r_x,0,r_z) &=0                      &\phiC(0,0,r_y,r_z) = 0&\\	
		\xi(Y,r_x,r_y,0) &=0                         &\xi(Y,0,0,r_z)= 0 &\\
                \xi(0,0,r_y,r_z) &=0\\
		\aver{\delta u \delta w}(Y,r_x,r_y,0) &= 0   &\aver{\delta u \delta w}(Y,0,0,r_z) = 0&\\
		\aver{\delta u \delta w}(0,0,r_y,r_z) &= 0 &
		\end{align*}
	\end{center}	

	\item  $\aver{\delta v \delta w}$

	\begin{center}
		\begin{align*}
		\phiR_x(Y,r_x,r_y,0) &= 0 \hspace{2cm}    &\phiR_x(0,r_x,0,r_z) =0&\\
		\phiR_x(0,0,r_y,r_z) &= 0 \\
		\phiR_y(Y,r_x,r_y,0) &= 0 \\
                \phiR_z(Y,0,0,r_z) &=0                    &\phiR_z(0,r_x,0,r_z) =0&\\
                \phiC(Y,r_x,r_y,0) &=0                    &\phiC(Y,0,0,r_z) =0&\\
                \phiC(0,0,r_y,r_z) &=0 \\
		\xi(Y,r_x,r_y,0) &= 0                        &\xi(Y,0,0,r_z)=0&\\
                \xi(0,r_x,0,r_z) &=0   \\	
		\aver{\delta v \delta w}(Y,r_x,r_y,0) &= 0   &\aver{\delta v \delta w}(Y,0,0,r_z) = 0&\\
                \aver{\delta v \delta w}(0,r_x,0,r_z) &= 0 &
		\end{align*}
	\end{center}	
\end{itemize}

\section{The ensemble-averaged quasi-streamwise vortex}
\label{sec:detection}

The procedure that yields the velocity field induced by the ensemble-averaged quasi-streamwise vortex used in \S\ref{sec:dudvqsv} is similar to that introduced by \cite{jeong-etal-1997}; the main steps of the procedure are described in the following.

The dominant vortical structure is educed from the present DNS database. Vortex candidates are searched first, defined as three-dimensional connected regions where the imaginary part $\lambda_{ci}$ of the complex conjugate eigenvalue pair of the velocity gradient tensor, also called swirling strength, exceeds the threshold $\lambda_{ci}^+>0.145$ \citep{zhou-etal-1999}. The connected region is built by assembling together 18-connected voxels \citep{rosenfeld-kak-1982}, i.e. voxels which are neighbors to every voxel that touches one of their faces or edges. 

Within each connected region, the centre of the vortex is defined as the point where $\lambda_{ci}$ is maximum; the orientation of the vortex axis is computed at the vortex centre. The orientation is given by the eigenvector associated with the real eigenvalue of the velocity gradient tensor \citep{chakraborty-balachandar-adrian-2005}. Vortices are then selected based on two additional criteria: i) their length in wall units must exceed 50 wall units, to exclude small structures in early or late stage of their life cycle \citep{jeong-etal-1997}; and (ii) their centre must be located within the region $21.2 \leq y^+ \leq 23.6$, the range of wall distances where several structures have been detected. The velocity fields of the selected quasi-streamwise vortices (approximately 14\% of all detected vortices) are eventually averaged together, by aligning all vortex centres together in the wall-normal plane and by accounting for the sense of rotation of the vortex, as given by the sign of the streamwise vorticity at the vortex centre.

\end{document}